\documentclass[%
 preprint, %linenumbers,
 amsmath,amssymb,
 aps, physrev,
]{revtex4-2}

\usepackage{graphicx}% Include figure files
\usepackage{dcolumn}% Align table columns on decimal point
\usepackage{bm}% bold math
\usepackage{amsmath}
\usepackage{mathtools}
\usepackage{todonotes}
\usepackage{comment}
\usepackage{caption}
\usepackage{subcaption}
\usepackage{tikz}
\usetikzlibrary{positioning}
\usetikzlibrary{calc}
\usetikzlibrary{matrix,fit,backgrounds}
\usepackage{xcolor}
\usepackage[most]{tcolorbox}
\colorlet{lightcyan}{cyan!40!}
\tcbset{on line, 
        boxsep=4pt, left=0pt,right=0pt,top=0pt,bottom=0pt,
        colframe=white,colback=lightcyan,  
        highlight math style={enhanced}
        }
\usepackage{lipsum}
\usepackage{float}

\newcommand{\fpp}[2]{\left (#1 \right )^{#2}}

\newif\ifcompletederivation
\completederivationfalse

\newif\iftohighlight
\tohighlighttrue
\newif\ifaddformulas
\addformulasfalse

\newif\ifstrictfloat
\strictfloatfalse

\begin{document}

\preprint{APS/123-QED}

%\title{\textbf{Grid-point and time-step requirements for simulations of two-phase flows.} 
\title{\textbf{Scalings and simulation requirements in two-phase flows} 
}% 

\author{Luis H. Hatashita}
\author{Pranav Nathan}%
\author{Suhas S. Jain}
 \email{Contact author: suhasjain@gatech.edu}
\affiliation{%
 Flow Physics and Computational Science Lab, \\
 %George W. Woodruff School of Mechanical Engineering, 
 Georgia Institute of Technology, Atlanta, GA, USA.
}%


\date{\today}% It is always \today, today,

\begin{abstract}

In this work, important two-phase flow scalings are derived, which enable the quantification of grid-point and time-step requirements as functions of Reynolds ($Re$), Weber ($We$), and Capillary ($Ca$) numbers. The adequate grid resolution is determined in the inertia-dominated regime with the aid of high-fidelity simulations of stationary two-phase homogeneous isotropic turbulence by evaluating convergence of total interfacial area, size distribution, Sauter mean diameter (SMD), and curvature distribution. Although standards for direct numerical simulations (DNS) for single-phase turbulence flow exist, there is a lack of similar guidance in two-phase flows. Therefore, length scale ratios of the Kolmogorov-Hinze to the Kolmogorov scale of $\eta_{KH}/\eta \sim We_{\mathcal{L}}^{-3/5}Re_{\mathcal{L}}^{3/4}$ in the inertia-dominated regime and the Kolmogorov-viscous to Kolmogorov scale of $\eta_{KV}/\eta \sim Ca_{\mathcal{L}}^{-1}Re_{\mathcal{L}}^{3/4}$ for the viscous-dominated regime, are constructed. These scalings imply a computational cost increase like $We_{\mathcal{L}}^{12/5}$ and $Ca_{\mathcal{L}}^4$, in the inertia-dominated and viscous-dominated regimes, respectively. A novel dimensionless number, coined as the \textit{ratio of interface scales} ($Ris$), defined as $Ris \coloneqq \eta_{KH}/\eta_{KV}=We_\mathcal{L}^{2/5}/Re_\mathcal{L}$, is proposed to aid in the classification of the turbulence regimes in the presence of an interface. 
Convergence of the total interfacial area, size distribution, SMD, and curvature distribution are observed for grid resolutions of $k_{\max} \eta_{KH} \geq 60$ for second-order schemes. Furthermore, it is observed that this lower bound is the minimum required to capture intermittent events responsible for the increase of instantaneous total interfacial area. 
This criterion will be a valuable tool for determining grid resolution and time-step requirements \textit{a-priori} for DNS of two-phase flows and for estimating the corresponding computational cost. This work provides guidelines and best practices for numerical simulations of two-phase flows, which will accelerate physics discovery and model development.

%\begin{description}
%\end{description}
\end{abstract}

%Use showkeys class option if keyword
\maketitle

%\tableofcontents

\section{Introduction}
\label{sec:intro}

% outline:
% - motivation - applications of multiphase (why it's important) and why numerical simulations are important;
% - true DNS doesn't exist (length scale can go to molecular level) - main challenge of numerical multiphase;
% - need instructions;
% - different than the single phase, where there's only kolmogorov, now have Hinze;
% - cite there are guidelines for single-phase, e.g., Choi and Moin, Yang, ;
% - therefore this is what we address in this paper

% writing starts here
% 1) context: 
%   a) motivation - ubiquity of multiphase;
High-fidelity numerical simulations, \textit{e.g.}, direct numerical simulations (DNS), of two-phase flows represent a viable alternative to experiments for the study of the interaction of wide-range of scale phenomena, such as turbulence with phase-separating surfaces undergoing drastic topological changes. One of the major implications of direct simulations of such bubble- and droplet-laden flows is the development of models for Reynolds-averaged Navier-Stokes (RANS) or large-eddy simulations (LES)~\cite{tryggvason:2013}. More accurate models would result in enhancing the predictive capabilities of numerical simulations that are important for various applications in nature and engineering, such as, breaking waves~\cite{chen:1999,deane:2002,melville1996role}, rain formation~\cite{ichikawa:2010,ghan:2011}, efficient design of bubble columns in chemical reactors~\cite{kantarci:2005}, boiling heat exchangers~\cite{dhir:1998}, atomization of liquid fuels~\cite{lin:1998,gorokhovski2008modeling}, to name a few.%\todo[]{citations for breaking waves, rain..}

%   b) limitations of simulations of two-phase, need for instructions
In DNS, the full range of spatial and temporal scales must be resolved on the grid in order to capture non-trivial interactions between the interface and all turbulence scales, and the resulting topological changes of the interface. On the other hand, in LES and RANS, the subgrid scales are modeled.
It maybe is impractical to perform a ``true DNS" of two-phase flows, since the smallest scales in the flow can reach molecular level during topological changes, e.g., breakup events. Nevertheless, it can be still referred to as DNS when the simulations resolve the most important scales in the flow and can faithfully capture the primary quantities of interest~\cite{tryggvason:2013}. 
Therefore there is a need for proper guidelines on the spatial and temporal resolutions, required for accurate numerical simulations, and the corresponding simulation cost estimates as a function of the relevant non-dimensional numbers, such as, Reynolds number ($Re$), Weber number ($We$), and Capillary number ($Ca$).
%how the cost would increase for higher Reynolds number, $Re$, and smaller bubbles/droplets (higher Weber, $We$, and higher Capillary, $Ca$, numbers, depending on the smallest diameters compare to the Kolmogorov length scale, $\eta$). 
%A concern with respect to the interface thickness scale not being actually resolved on the grid may be raised, due to being on the order of nanometers compared to smallest turbulent scales on the order of the micrometers~\cite{}. Nevertheless, one may argue that interface-capturing and -tracking methods solved in conjunction with the Navier-Stokes equation within the whole range of scales without any turbulence modeling still provide accurate results compared to experiments and it is also able to properly capture phenomena such as breakup and coalescence~\cite{tryggvason:2013}. Therefore, not limiting the scope of this work.

% 3) grid point and time step papers review
\subsection{Resolution requirements for numerical simulations} 
\label{subsec:res_req_intro}

One of the first studies to provide estimates of the required number of grid points for LES of turbulent boundary layers, relevant for aerospace applications, was by Chapman in 1979~\cite{chapman:1979}. This work proposed the use of nested grids and used the scaling relation of boundary layer thickness ($\delta$) with $Re_x = Ux/\nu$ to compute the grid estimates of $N_{wm}\sim Re_{L_x}^{2/5}$ for wall-modeled LES (WMLES) and $N_{wr} \sim Re_{L_x}^{9/5}$ for wall-resolved LES (WRLES), based on the flat-plate length in the streamwise direction $L_x$. Choi and Moin (2012)~\cite{choi:2012} improved upon these estimates by updating the relations for $\delta$ and skin friction coefficient $c_f$ to more realistic $Re_x$, and integrating $\delta$ over the streamwise direction instead of using an average value proposed by Chapman. Their proposed estimates are $N_{wm}\sim Re_{L_x}$, $N_{wr} \sim Re_{L_x}^{13/7}$, and $N_{DNS} \sim Re_{L_x}^{37/14}$, for WMLES, WRLES, and DNS, respectively. Yang and Griffin (2021)~\cite{yang:2021a} were able to relax the scalings of \cite{choi:2012} by introducing the concept of nested grids in all directions, improving the DNS scaling to $N_{DNS} \sim Re_{L_x}^{115/56}$. Furthermore, they also provided guidelines for time-step requirements, which are directly related to the total computational cost of the simulations. The total number of time steps would scale as $N_{t,\rm WMLES} \sim Re_{L_x}^{1/7}$, $N_{t,\rm WRLES} \sim Re_{L_x}^{6/7}$, and $N_{t,\rm DNS} \sim Re_{L_x}^{6/7}$. 

In addition to the grid count scalings with $Re$, the spatial grid resolution is also an important quantity that determines the total grid count and the cost of the numerical simulations. A customary grid spacing resolution of $k_{max} \eta \approx 1.5$ for turbulence~\cite{yeung:2018} and $\Delta y^+ \lesssim 1$ near walls for wall-bounded flows~\cite{kim:1987} have been shown to be sufficient to capture low-order statistics, where $\eta$ is the Kolmogorov scale, $k_{max}$ is the maximum resolvable wave number, and, $\Delta y^+$ is the wall-normal grid spacing in the inner units (normalized by the viscous length scale $\nu/u_\tau = \nu / \sqrt{\tau|_{\rm wall} / \rho}$). However, for accurate prediction of extreme events of energy dissipation rate, enstrophy, and wall shear stress, higher resolutions may be required ~\cite{yeung:2018,yang:2021b}.  
%may be jeopardized. Therefore, Yeung \textit{et al.} (2018)~\cite{yeung:2018} proposed a study to evaluate this discrepancy both in spatial and temporal resolutions for high $Re_\lambda$. Their results indeed showed that not sufficiently resolved grid spacing yielded in narrower PDFs of dissipation and enstrophy, and large magnitude events showed significant sensitivity to time increment. 
%Yang \textit{et al.} (2021)~\cite{yang:2021b} extended the previous study for wall-bounded flows, where the prediction of extreme shear stress events is relevant for properly capturing particle entrainment.

These grid count estimates and resolution requirements are all defined in the context of single-phase turbulence and wall-bounded flows. However, such guidelines are lacking in two-phase flows, where the presence of the interface with surface tension effects can introduce additional length and time scales to be resolved in the numerical simulations. 

% 2) physics of two-phase
\subsection{Scalings in two-phase flows}
\label{subsec:scalings_intro}

In single-phase turbulence, the smallest length scale is the Kolmogorov scale ($\eta$), which is determined by the fluid viscosity and energy dissipation rate. Nevertheless, for sufficiently high $Re$ in two-phase flows, there exists an additional length scale, known as the Kolmogorov-Hinze scale ($\eta_{KH}$) \cite{kolmogorov:1949,hinze:1955}, below which the turbulence is no longer strong enough to overcome surface tension and hence ceasing the breakup of bubbles and drops .
%introduces the interaction of surface tension with the flow field. Breakup essentially ceases at a scale, $\eta_{KH}$, first defined by Kolmogorov (1949)~\cite{kolmogorov:1949} and Hinze (1952)~\cite{hinze:1955}, since turbulence is no longer strong enough to overcome surface tension. 
%A restricting assumption of the Kolmogorov-Hinze framework is that the bubble/drop diameter is in the inertial range (\textit{i.e.}, $\eta_{KH} > \eta$)
The inclusion of this additional length scale imposes an extra set of constraints with respect to high-fidelity numerical simulations. For increasing $Re_{\ell}=u'\ell/\nu$, the range of viscous length scales increases as $\ell/\eta \sim Re_\ell^{3/4}$~\cite{pope:2000}, thus requiring grid spacing to be reduced, increasing the computational cost and demand for computer memory. In two-phase turbulence, as it will be shown later in this work, for increasing $We_\ell = \rho u'^2 \ell / \sigma$, the range of interface length scales increases as $\ell/\eta_{KH} \sim We_\ell^{3/5}$, this means that as the surface tension is reduced (\textit{e.g.}, smaller surface tension coefficient $\sigma$), smaller and smaller bubbles/droplets are generated, imposing harsher constraints to accurately capture the dispersed phase. When $We_\ell$ is increased to a very high value, the bubbles/droplets become small enough that the viscous effects dominate and balance the surface tension instead of the inertia from the turbulence \cite{vankova:2007,ni:2024}. A new length scale emerges in this regime, known as the Kolmogorov-viscous scale ($\eta_{KV}$). Hence, there is a need for a systematic study on scalings in terms of $Re$, $We$, and $Ca$, in different regimes. 

%\todo[]{add the schmatic from your presentation here}
%\todo[inline]{comment figure}

Figure~\ref{fig:scales_turb} illustrates the different reference length scales in single-phase, two-phase inertia-dominated, and viscous-dominated turbulence regimes. For the first, the limiting scale is only the Kolmogorov scale $\eta$, where viscosity is strong enough to dissipate kinetic energy into internal energy. For the latter two, the presence of surface tension can introduce additional limiting scales. Depending on the combination of non-dimensional numbers, the interface scales may lie in the inertial or viscous regimes, respectively. In the inertial regime, the nature of bubbles/droplets can be distinguished by the Kolmogorov-Hinze $\eta_{KH}$ scale, scales below which they are mostly spherical due to surface energy minimization and above which they can be highly corrugated/distorted. In the viscous regime, viscous stresses are likely to mostly deform the interface. 

\ifstrictfloat
\begin{figure}[H]
\else
\begin{figure}
\fi
    \centering
   % \begin{subfigure}[c]{0.32\linewidth}
   %     %\captionsetup{
   %     %  justification=raggedright,
   %     %  singlelinecheck=false,
   %     %  font=small,
   %     %  position=top % ensures alignment by top of caption box
   %     %}
   %     %(a)
   %     \includegraphics[width=0.5\linewidth]{figs/schem/single-phase.pdf}
   %     %\includegraphics[height=0.2\textheight]{figs/schem/single-phase.pdf}
   %     \caption{}
   % \end{subfigure}
   % \begin{subfigure}[c]{0.32\linewidth}
   %     %\captionsetup{
   %     %  justification=raggedright,
   %     %  singlelinecheck=false,
   %     %  font=small,
   %     %  position=top % ensures alignment by top of caption box
   %     %}
   %     %(b)
   %     \includegraphics[width=0.85\linewidth]{figs/schem/two-phase-inertial.pdf}
   %     %\includegraphics[height=0.2\textheight]{figs/schem/two-phase-inertial.pdf}
   %     \caption{}
   % \end{subfigure}
   % \begin{subfigure}[c]{0.32\linewidth}
   %     %\captionsetup{
   %     %  justification=raggedright,
   %     %  singlelinecheck=false,
   %     %  font=smalk,
   %     %  position=top % ensures alignment by top of caption box
   %     %}
   %     %(c)
   %     \includegraphics[width=0.6\linewidth]{figs/schem/two-phase-viscous.pdf}
   %     %\includegraphics[height=0.2\textheight]{figs/schem/two-phase-viscous.pdf}
   %     \caption{}
   % \end{subfigure}
    \includegraphics[width=\textwidth]{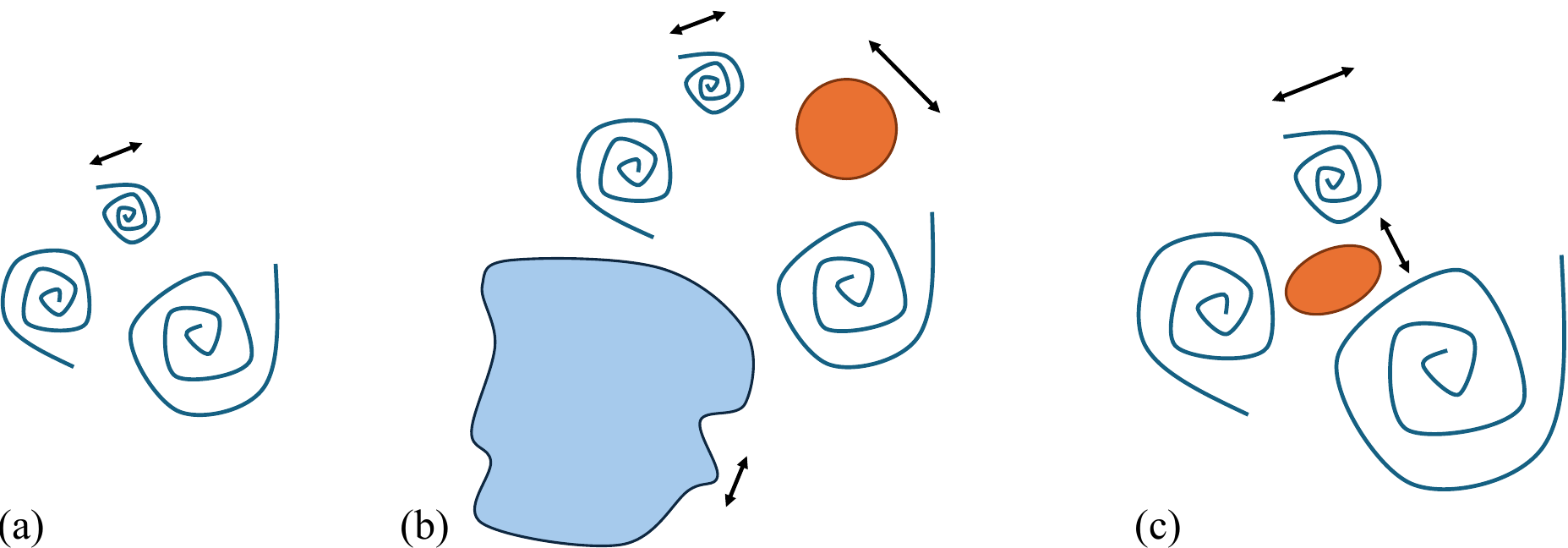}
    \caption{Scales of interest in single-phase (a), two-phase inertia-dominated (b), and two-phase viscous-dominated turbulence (c). Reference length scales highlighted by black arrows. In the inertia-dominated regime (b), super-Kolmogorov-Hinze scale is denoted by the blue blob and the sub-Kolmogorov-Hinze by the orange sphere. In the viscous-dominated regime (c), the orange ellipsoid represents the Kolmogorov-viscous scale deformed by viscous stresses below the Kolmogorov scale.}%\todo[]{It might be better to add (a), (b), and (c) instead. Its not very formal to use left and right in journal papers.}
    \label{fig:scales_turb}
\end{figure}

%\todo[]{need to cite Rui Ni paper somewhere..}

\subsection{Quantities of interest in two-phase flows}
% 4) quantities of interest
%\todo[inline]{explain briefly quantities of interest in two-phase and how they could be related to low or high-order statistics, justifying our study}
The main quantities of interest for two-phase flows targeted in this work are total interfacial area, size distribution, Sauter mean diameter (SMD), and curvature distribution. 
%A brief description of each and their follows. 
Interfacial area is an important quantity that governs mass, momentum, and energy transfer across the interface, \textit{e.g.}, mass transfer during vaporization~\cite{schlottke:2008,hardt:2008,soh:2016} and heat transfer in boiling~\cite{dhir:1998}. 
It is important to accurately capture both the local and total interfacial area, otherwise mass and heat transfer predictions will be incorrect. 
Because of the importance of this quantity, both physics-based models \cite{hatashita:2025} and phenomenological ones \cite{vallet:2001,lebas:2009,chesnel:2011,granger:2024} have been developed in the context of RANS and LES. The proposed estimates for high-fidelity simulations of two-phase flows would further aid in the development of these models in the future.
%that are physics-based~\cite{hatashita:2025} or by transporting it via another equation~. 
%Irrespective of the approach used, high-fidelity simulation data is required to accurately capture local and total interfacial area, otherwise mass and heat transfer predictions will be incorrect. 

Particles, sediments, bubbles, and droplets are statistically described via size distributions, $n(d)$ where $n$ is the number of bubbles/droplets of diameter $d$, even for atomizers that can produce the narrowest of the distributions~\cite{williams:1990,kowalczuk:2016}. Furthermore, based on the size distributions, one may compute the total interfacial area as $A = \sum_{d_{\min}}^{d_{\max}} \pi d^2 n(d)$. Therefore, size distributions can also be used to quantify mass and heat transfer. Full-scale simulations of combustors oftentimes take experimentally-obtained distributions as boundary conditions of fuel injectors when using Lagrangian-tracked droplets~\cite{apte:2003,senoner:2009}. These types of simulations often neglect primary breakup regions and inject the distribution of fully developed sprays, possibly missing interactions within this zone. High-fidelity simulations should aim for resolving both regions. This work aims to provide resolution requirements for capturing both regimes, both the early stage primary breakup dominated by interfacial corrugations and the later stage dominated by dispersed phase.
%primary breakup zones at the early stages of forcing, and dilute/dispersed zones when the stationary state is achieved and the size distribution is converged. 

A general expression for a diameter based on the distribution $n(d)$ can be defined as
\begin{equation}
\label{eq:d_equation}
d_{pq} = \frac{\sum  d^p n(d)}{\sum d^q n(d)},
\end{equation}
for instance, SMD is computed with $p=3$ and $q=2$, $d_{32}$. It is known as the volume-to-surface ratio diameter, since the total volume scales with $d^3$ and total surface with $d^2$~\cite{kowalczuk:2016}. It is commonly used as a quantity to compute average combustion or heat transfer characteristics, hence predicting performance~\cite{estes:1995,semiao:1996,pacek:1998}. It is noteworthy that although estimates may be obtained via average quantities such as SMD, with large-scale systems, size distributions vary in space, so does SMD~\cite{williams:1990}. 

Accurate predictions of the transient state between liquid injection and fully broken spray cannot be described only using size distributions, since primary and secondary breakup are mostly characterized by corrugations and non-spherical structures, yielding size distributions meaningless. Curvature distributions were proposed to describe length scales associated with initial breakup, precursor to final size distributions~\cite{zandian:2019,goodrich:2022,thakkar:2024}. Furthermore, in emulsions, curvature has been suggested to be one of the physical indicators of phase inversion~\cite{dickinson:1981,strey:1996,vikhansky:2020}, which is a low-energy method for producing fine emulsions~\cite{kumar:2015}.

% 5) objectives of the paper
Therefore, the main objectives of this paper are to (a) analyze and derive relevant physical length scales in two-phase turbulence in different regimes and how they vary with non-dimensional numbers; (b) use the scalings to derive estimates of grid-point and time-step requirements for high-fidelity simulations; and (c) evaluate and verify spatial grid resolution requirements using numerical simulations based on the primary quantities of interest, such as, total interfacial area, size distributions, SMD, and curvature distributions. The remainder of the paper is structured as follows: in the Section~\ref{sec:scalings}, scalings for the ratio of interface to turbulence length scales are derived; in the Section~\ref{sec:reqs}, based on the scalings, grid-point and time-step requirements are derived as functions of spatial and temporal resolutions; in the Section~\ref{sec:verify}, the convergence of the two-phase quantities of interest are verified in stationary homogeneous isotropic turbulence (HIT), demonstrating resolution requirements for the numerical methods adopted herein; lastly in the Section~\ref{sec:conc}, 
%a brief discussion is posed on how these results should be extended to other numerical methods. 
final remarks are presented.

\section{Scalings in two-phase flows}
\label{sec:scalings}

% - derive length scale ratios (scalings) in two-phase flows
% - Hinze in terms of large scale quantities, Re_l and We_l, or Re_lambda and We_lambda, or in terms of k, \epsilon

% - talk about viscous dominated regime;
% - will need capillary number dependence;
% - put phase plot defining all regions;

% writing starts here
% 1) derivation of scalings, both combinations in terms of l and lambda
%\todo[inline]{do i need to consider phasic $Re$ and $We$?}
%\todo[inline]{If one uses only $Ca$, the scaling is straight forward, just need to be aware if $\mu_d \gg \mu_c$ such that $Oh \gg 1$, becomes a whole other mess.}
%\todo[inline]{highlight assumptions of unity density and viscous ratio}

In this section, first, the ranges of scales in two-phase turbulence are derived for both inertia- and viscous-dominated regimes in Section~\ref{subsec:scaling_inertial} and Section~\ref{subsec:scaling_viscous}. Second, a discussion on the ratio of inertia- and viscous-dominated limiting scales is presented in Section~\ref{subsec:scaling_ratio}, along with a new non-dimensional number. 

\subsection{Inertia-dominated regime}
\label{subsec:scaling_inertial}

In what follows, the range of interface to turbulence scales in the inertia-dominated regime, $\eta_{KH}/\eta$, is constructed. This requires defining reference scales, such as energy-containing large scale, $\mathcal{L}$, Taylor microscale, $\lambda$, Kolmogorov scale, $\eta$, and Kolmogorov-Hinze scale, $\eta_{KH}$. Dimensionless numbers take reference scales as arguments, hence, yielding in explicit expressions for those scales in terms of Reynolds, $Re$, and Weber, $We$, numbers. Based on the explicit expressions of the scales, $l_{\rm ref} = l_{\rm ref} (Re, We)$, their ratios are obtained by division and algebraic manipulation, resulting in power law functions of $Re$ and $We$.

In single-phase turbulence, the reference length scales are: the large scale representing the energy-containing eddies, $\mathcal{L}=k^{3/2}/\epsilon$, where $k$ is the turbulent kinetic energy and $\epsilon$ the energy dissipation rate; the Taylor microscale, $\lambda$; and the Kolmogorov scale, $\eta=(\nu^3/\epsilon)^{1/4}$, where $\nu$ is the kinematic viscosity~\cite{pope:2000}. The corresponding Reynolds numbers are: $Re_{\mathcal{L}} = k^{1/2}\mathcal{L}/\nu$; $Re_\lambda = u'\lambda/\nu$ (where $u'$ is the root-mean-squared velocity fluctuation); and $Re_\eta = u'\eta/\nu= O(1)$. Therefore, we may construct the range of viscous length scales, which provides an estimate for the grid requirements, since it characterizes the scaling of the so-dwindling small scales, still to be resolved by the computational grid points as $Re$ increases.
\begin{equation}
    \label{eq:eta_l}
    \frac{\eta}{\mathcal{L}} \sim Re_\mathcal{L}^{-3/4} = \left (\frac{3}{20} \right )^{-3/4} Re_\lambda^{-3/2}.
\end{equation}

In two-phase turbulence, the presence of an interface introduces the effects of surface tension on the flow. It is described by the Weber number, for the large-scale, define its Weber number, $We_\mathcal{L}$, as~\cite{hinze:1955}
\begin{equation}
    \label{eq:we_l}
    We_\mathcal{L} = \frac{\rho u'^2 \mathcal{L}}{\sigma},
\end{equation}
which measures the ratio between the inertial, $\rho u'^2$, and surface tension, $\sigma / \mathcal{L}$ forces (where $\sigma$ is the surface tension coefficient). 
\begin{comment}
The large scale Weber number can then be written in terms of $k$ and $\epsilon$, 
%as a function of the energy cascade quantities follow from the substitution of 
%using Eq.~\eqref{eq:scale_int} and $k = 3u'^2/2$,
%in the Eq.~\eqref{eq:we_l}, 
as
%
\ifcompletederivation
\begin{equation}
    \label{eq:we_l_2} 
    We_\mathcal{L} = \frac{\rho}{\sigma} \frac{2k}{3} \frac{k^{3/2}}{\epsilon} = \frac{2}{3} \frac{\rho k^{5/2}}{\sigma \epsilon}.
\end{equation}
\else
\begin{equation}
    \label{eq:we_l_2} 
    We_\mathcal{L} = %\frac{\rho}{\sigma} \frac{2k}{3} \frac{k^{3/2}}{\epsilon} = 
    \frac{2}{3} \frac{\rho k^{5/2}}{\sigma \epsilon}.
\end{equation}
\fi
\end{comment}

The comparison of turbulence and interface scales requires the definition of a characteristic interface length scale to be compared with the Kolmogorov scale, $\eta$. 
The turbulent Weber number is defined as 
\begin{equation}
    \label{eq:we_t} 
    We_t = \frac{\rho u_D^2 D}{\sigma},
    %We_t^c = \frac{\rho u_D^2 D}{\sigma},
\end{equation}
where $u_D$ is the first-order structure function corresponding to the length scale $D$. By the Kolmogorov (1941)~\cite{kolmogorov:1941} theory, one can express $u_D^2 \sim (\epsilon D)^{2/3}$ in the inertial range. 
Furthermore, from the Kolmogorov-Hinze theory~\cite{kolmogorov:1949,hinze:1955}, further assuming homogeneous and isotropic turbulence and the bubble/drop diameter to be within the inertial range, \textit{i.e.}, $\eta \ll D \ll \mathcal{L}$, breakup is hypothesized to be driven by the dynamic pressure fluctuations caused by changes in velocity over distances at most equal to the diameter $D$. Therefore, it is expected to stop at a specific scale, denoted as the Kolmogorov-Hinze scale, $\eta_{KH}$ with $We_t = O(1)$, similarly defined as in single-phase at an order-of-unity dimensionless number, \textit{e.g.}, $Re_\eta = O(1)$, $We_{\eta_{KH}} = We_t^c = O(1)$. 
%Herein, the critical turbulent Weber number, $We_t^c$, characterizes this demarcation of limiting breakup, where 
Thus, we can obtain an expression for the diameter of the smallest bubble/droplets in the inertial-range of turbulence, $D^c$, (also referred to as $\eta_{KH}$) as
\ifcompletederivation
\begin{eqnarray}
    \label{eq:eta_kh}
    & We_t^c & \sim \frac{\rho}{\sigma} (\epsilon D^c)^{2/3} D^c = \frac{\rho}{\sigma} \frac{3}{2} \frac{u'^2}{l^{2/3}} {D^c}^{5/3}, \nonumber \\
    \therefore \; & D^c & \sim \left (\frac{We_t^c \sigma}{\rho u'^2} l^{2/3} \right )^{3/5}.
\end{eqnarray}
\else
\begin{equation}
    \label{eq:eta_kh}
    We_t^c \sim \frac{\rho}{\sigma} (\epsilon D^c)^{2/3} D^c% = \frac{\rho}{\sigma} \frac{3}{2} \frac{u'^2}{l^{2/3}} {D^c}^{5/3}, \nonumber \\
    \; \therefore \; D^c \sim \left (\frac{We_t^c \sigma}{\rho u'^2} \mathcal{L}^{2/3} \right )^{3/5}.
\end{equation}
\fi
Several studies have focused on determining the value of $We_t^c$~\cite{ni:2024}. For reference, experiments in microgravity by Risso and Fabre (1998)~\cite{risso:1998} reported $We_t^c \approx 4.5$ and lattice Boltzman simulations of HIT in a 3D periodic box by Qian \textit{et al.} (2006)~\cite{qian:2006} yielded a value of approximately $3$.

The range of interface length scales similarly imposes an additional constraint on the grid requirements if the smallest bubbles and droplets are to be accurately resolved on the computational grid as $We$ increases. %Therefore, without loss of generality, one may take $We_t^c = 1 = \mathcal{O}(1)$ and 
Rewriting Eq.~\eqref{eq:eta_kh} as a ratio of the Kolmogorov-Hinze to the large length scales, we obtain
\ifcompletederivation
\begin{eqnarray}
    \label{eq:eta_kh_l1}
    \frac{\eta_{KH}}{l} & \sim & \left (We_t^c \frac{\sigma}{\rho u'^2 l} \right )^{3/5} = \left (\frac{We_t^c}{We_l} \right )^{3/5}, \\
    \label{eq:eta_kh_l2}
    & \sim & \left (We_t^c \frac{\sigma}{\rho u'^2 \lambda} \frac{\lambda}{l} \right )^{3/5} \sim \left (\frac{We_t^c}{We_\lambda Re_\lambda} \right)^{3/5}.
\end{eqnarray}
\else
\begin{equation}
    \label{eq:eta_kh_l}
    \frac{\eta_{KH}}{\mathcal{L}} \sim \left (We_t^c \frac{\sigma}{\rho u'^2} \mathcal{L}^{2/3} \right )^{3/5} \frac{1}{\mathcal{L}} = \left (\frac{We_t^c}{We_\mathcal{L}} \right )^{3/5}
    \sim \left (\frac{We_t^c}{We_\lambda Re_\lambda} \right)^{3/5}.
\end{equation}
\fi
Lastly, the relative ratio of interface and viscous scales can be constructed using Eqs.~\eqref{eq:eta_l} and~\eqref{eq:eta_kh_l} as%~\ref{eq:eta_kh_l1},~\ref{eq:eta_kh_l2}:
\ifcompletederivation
\begin{eqnarray}
    \label{eq:eta_kh_eta}
    \frac{\eta_{KH}}{\eta} = \frac{\eta_{KH}}{\mathcal{L}} \frac{\mathcal{L}}{\eta} & \sim & \left (\frac{We_t^c}{We_\mathcal{L}}\right )^{3/5} Re_\mathcal{L}^{3/4}, \\
    & \sim & \left (\frac{We_t^c}{We_\lambda Re_\lambda} \right )^{3/5} Re_\lambda^{3/2} = \left (\frac{We_t^c}{We_\lambda} \right )^{3/5} Re_\lambda^{9/10}.
\end{eqnarray}
\else
\begin{equation}
    \label{eq:eta_kh_eta}
    \frac{\eta_{KH}}{\eta} = \frac{\eta_{KH}}{\mathcal{L}} \frac{\mathcal{L}}{\eta} \sim \left (\frac{We_t^c}{We_\mathcal{L}}\right )^{3/5} Re_\mathcal{L}^{3/4} 
    \sim \left (\frac{We_t^c}{We_\lambda} \right )^{3/5} Re_\lambda^{9/10}.
\end{equation}
\fi

\subsection{Viscous-dominated regime}
\label{subsec:scaling_viscous}
% - talk about viscous dominated regime;
% - will need capillary number dependence;

%\todo[inline]{If one uses only $Ca$, the scaling is straight forward, just need to be aware if $\mu_d \gg \mu_c$ such that $Oh \gg 1$, becomes a whole other mess.}

For sufficiently weak surface tension, the length scale, at which breakup will cease, can be comparable or even smaller than the Kolmogorov length scale. In this regime, namely the viscous-dominated regime, breakup occurs under the action of viscous stresses~\cite{vankova:2007,ni:2024}. Analogous to the turbulent Weber number, $We_t$, there exists a turbulent Capillary number, $Ca_t$, which describes the strength of viscous stresses compared to surface tension. There is also a corresponding critical Capillary number, $Ca_t^c$, which denotes the point of limiting breakup. First, define the turbulent Capillary number as
\begin{equation}
    \label{eq:ca_t}
    Ca_t = \frac{\sqrt{\rho \mu \epsilon} D}{\sigma}.
\end{equation}
The corresponding critical diameter (also referred to as Kolmogorov-viscous scale, $\eta_{KV}$) in the viscous-dominated regime is
\begin{equation}
    \label{eq:eta_kv}
    D^c = \frac{Ca_t^c \sigma}{\sqrt{\rho \mu \epsilon}}.
\end{equation}
The range of interface length scales in the viscous regime imposes harsher resolution constraints for increasing $Ca$. %Rearrange Eq.~\ref{eq:eta_kv} as a ratio of the interface to the integral length scales,
Constructing the ratio of Kolmogorov-viscous scale to the large turbulent scales, we get
\ifcompletederivation
\begin{eqnarray}
    \label{eq:eta_kv_l1}
    \frac{\eta_{KV}}{\mathcal{L}} & = & Ca_t^c \frac{\sigma}{\sqrt{\rho \mu \epsilon}} \frac{1}{\mathcal{L}} = \frac{Ca_t^c}{Ca_\mathcal{L}}, \\
    \label{eq:eta_kv_l2}
    & = & Ca_t^c \frac{\sigma}{\sqrt{\rho \mu \epsilon} \lambda} \frac{\lambda}{\mathcal{L}} \sim \frac{Ca_t^c}{Ca_\lambda Re_\lambda}.
\end{eqnarray}
\else
\begin{equation}
    \label{eq:eta_kv_l}
    \frac{\eta_{KV}}{\mathcal{L}} = Ca_t^c \frac{\sigma}{\sqrt{\rho \mu \epsilon}} \frac{1}{\mathcal{L}} = \frac{Ca_t^c}{Ca_\mathcal{L}} \sim \frac{Ca_t^c}{Ca_\lambda Re_\lambda}.
\end{equation}
\fi
The ratio between the Kolmogorov-viscous scale and viscous scales can be constructed using Eqs.~\eqref{eq:eta_l} and~\eqref{eq:eta_kv_l} as%~\ref{eq:eta_kv_l1},~\ref{eq:eta_kv_l2}:
\ifcompletederivation
\begin{eqnarray}
    \label{eq:eta_kv_eta}
    \frac{\eta_{KV}}{\eta} = \frac{\eta_{KV}}{\mathcal{L}} \frac{\mathcal{L}}{\eta} & \sim & \frac{Ca_t^c}{Ca_\mathcal{L}} Re_\mathcal{L}^{3/4}, \\
    & \sim & \frac{Ca_t^c}{Ca_\lambda Re_\lambda} Re_\lambda^{3/2} = \frac{Ca_t^c}{Ca_\lambda} Re_\lambda^{1/2} .
\end{eqnarray}
\else
\begin{equation}
    \label{eq:eta_kv_eta}
    \frac{\eta_{KV}}{\eta} = \frac{\eta_{KV}}{\mathcal{L}} \frac{\mathcal{L}}{\eta} \sim \frac{Ca_t^c}{Ca_\mathcal{L}} Re_\mathcal{L}^{3/4} \sim \frac{Ca_t^c}{Ca_\lambda} Re_\lambda^{1/2} .
\end{equation}
\fi

It is important to mention that the Capillary number is sufficient to describe length scales when the dispersed phase viscosity is on the order of or smaller than the carrier phase viscosity. When the viscosity of the dispersed phase is significantly larger than the viscosity of the carrier phase, the dispersed phase viscosity can act as a damping force, resisting the deformation. In this regime, Ohnesorge number, $Oh$, also needs to be considered in the analysis. 
%Herein as an initial effort, the focus is limited to unity and small enough viscosity ratios, $\mu_d / \mu_c \ll 1$.

%\subsection{A new non-dimensional number}
\subsection{On the characterization of interface scales in turbulence}
\label{subsec:scaling_ratio}

%\todo[inline]{does it make sense to add $\eta_{KH}/\eta_{KV}$?}

Although we may calculate the range of inertial ($\eta_{KH}$) and viscous ($\eta_{KV}$) interface scales with respect to a single-phase reference, such as the large or Kolmogorov length scales, there is no explicit way to identify the interfacial flow regime using these length scale ratios, \textit{i.e.}, is the critical length scale in the viscous dissipative regime or inertial range of turbulence? Therefore, an expression for the turbulent-inertial to turbulent-viscous ratio of the interface scales, $\eta_{KH}/\eta_{KV}$, would be of interest, which is analogous to the role of the Reynolds number in single-phase turbulence. Now, taking the ratio of the Eqs.~\eqref{eq:eta_kh_l} and~\eqref{eq:eta_kv_l}, we get
\ifcompletederivation
\begin{eqnarray}
    \label{eq:eta_kh_eta_kv}
    \frac{\eta_{KH}}{\eta_{KV}} = \frac{\eta_{KH}}{l} \frac{l}{\eta_{KV}} & \sim & \left (\frac{We_t^c}{We_l} \right )^{3/5} \frac{Ca_l}{Ca_t^c} \\
    & \sim & \left (\frac{We_t^c}{We_\lambda Re_\lambda} \right )^{3/5} \frac{Ca_\lambda Re_\lambda}{Ca_t^c} = \left (\frac{We_t^c}{We_\lambda} \right )^{3/5} \frac{Ca_\lambda}{Ca_t^c} Re_\lambda^{2/5}.
\end{eqnarray}
\else
\begin{equation}
    \label{eq:eta_kh_eta_kv}
    \frac{\eta_{KH}}{\eta_{KV}} = \frac{\eta_{KH}}{\mathcal{L}} \frac{\mathcal{L}}{\eta_{KV}} \sim \left (\frac{We_t^c}{We_\mathcal{L}} \right )^{3/5} \frac{Ca_\mathcal{L}}{Ca_t^c} 
    \sim \left (\frac{We_t^c}{We_\lambda} \right )^{3/5} \frac{Ca_\lambda}{Ca_t^c} Re_\lambda^{2/5}.
\end{equation}
\fi
We can construct an expression for the \textit{ratio of interface scales}, $Ris$, (a new non-dimensional number) in turbulence, as% is related to the Ohnesorge number, $Oh$, as follows
\ifcompletederivation
\begin{eqnarray}
    \label{eq:subst_eta_kh_eta_kv}
    \frac{\eta_{KH}}{\eta_{KV}} & \sim & \frac{\mu u'}{\sigma} \left ( \frac{\sigma}{\rho u'^2 l} \right )^{3/5} \nonumber \\
    & \sim & \left (\frac{We_l^2}{Re_l^{5}}  \right )^{1/5} = \frac{We_l^{2/5}}{Re_l} = Nd
\end{eqnarray}
\else
\begin{equation}
    \label{eq:subst_eta_kh_eta_kv}
    \frac{\eta_{KH}}{\eta_{KV}} \sim \frac{\mu u'}{\sigma} \left ( \frac{\sigma}{\rho u'^2 \mathcal{L}} \right )^{3/5}
    \sim \iftohighlight
    \tcbhighmath{\frac{We_\mathcal{L}^{2/5}}{Re_\mathcal{L}} \eqcolon Ris.}
    \else
    \frac{We_\mathcal{L}^{2/5}}{Re_\mathcal{L}} \eqcolon Ris.
    \fi
\end{equation}
\fi
If we define the Ohnesorge number, $Oh$, as
\begin{equation}
    \label{eq:oh}
    Oh_l = \frac{\mu}{\sqrt{\rho \sigma l}} = \frac{\sqrt{We_l}}{Re_l},
\end{equation}
then the new non-dimensional number
%If one defines the ratio of interface scales in turbulence as a non-dimensional number, 
$Ris = We_\mathcal{L}^{2/5}/Re_\mathcal{L}$, can be related to the Ohnesorge number as follows
%Therefore, Eq.~\ref{eq:subst_eta_kh_eta_kv} can be rearranged as
%
\ifcompletederivation
\begin{eqnarray}
    \label{eq:relation_ris_oh} 
    Ris^{5/4} & = & \frac{We_l^{1/2}}{Re_l^{5/4}} = \frac{Oh}{Re_l^{1/4}} \nonumber \\
    \therefore Ris & = & \left ( \frac{Oh_l^{4}}{Re_l} \right )^{1/5}
\end{eqnarray}
\else
\begin{equation}
    \label{eq:relation_ris_oh} 
    Ris^{5/4} = \frac{We_\mathcal{L}^{1/2}}{Re_\mathcal{L}^{5/4}} = \frac{Oh_\mathcal{L}}{Re_\mathcal{L}^{1/4}}
    \; \therefore \; 
    \iftohighlight
    \tcbhighmath{Ris = \left ( \frac{Oh_\mathcal{L}^{4}}{Re_\mathcal{L}} \right )^{1/5}.}
    \else
    Ris = \left ( \frac{Oh_\mathcal{L}^{4}}{Re_\mathcal{L}} \right )^{1/5}.
    \fi
\end{equation}
\fi

% - put phase plot defining all regions;

The $Ris$ number can be interpreted as an extension of the $Oh$ number in the presence of turbulence. The Table~\ref{tab:summary_multi_turb_scales} summarizes the power law scalings of the relevant length scale ratios in two-phase turbulence as functions of $Re$, $We$, and $Ca$.
\ifstrictfloat
\begin{table}[H]%% placement specifier
\else
\begin{table}
\fi
\centering%% For centre alignment of tabular.
\caption{Power laws for relevant scales in single-phase turbulence, two-phase inertia-dominated, and viscous-dominated interface regimes as functions of $Re$, $We$, and $Ca$.}\label{tab:summary_multi_turb_scales}
\begin{tabular}{c | c | c}%% Table column specifiers
  \hline
   Ratio                   & Large scale                                 & Taylor scale \\ \hline
   $\eta/\mathcal{L}$      & $Re_\mathcal{L}^{-3/4}$                     & $Re_\lambda^{-3/2}$    \\
   $\eta_{KH}/\mathcal{L}$ & $We_\mathcal{L}^{-3/5}$                     & $We_\lambda^{-3/5}Re_\lambda^{-3/5}$    \\
   $\eta_{KH}/\eta$        & $We_\mathcal{L}^{-3/5}Re_\mathcal{L}^{3/4}$ & $We_\lambda^{-3/5}Re_\lambda^{9/10}$    \\
   $\eta_{KV}/\mathcal{L}$ & $Ca_\mathcal{L}^{-1}$                       & $Ca_\lambda^{-1}Re_\lambda^{-1}$        \\
   $\eta_{KV}/\eta$        & $Ca_\mathcal{L}^{-1}Re_\mathcal{L}^{3/4}$   & $Ca_\lambda^{-1}Re_\lambda^{1/2}$       \\
   $\eta_{KH}/\eta_{KV}$   & $We_\mathcal{L}^{-3/5}Ca_\mathcal{L}$       & $We_\lambda^{-3/5}Ca_\lambda Re_\lambda^{2/5}$ \\ \hline
\end{tabular}
\end{table}

Figure~\ref{fig:regimes_interface} summarizes the different regimes of interface scales in turbulence. In the inertia-dominated regime in Fig.~\ref{fig:regimes_interface} (a), $\eta_{KH} \gg \eta$, for increasing $Re_\lambda$, the $\eta_{KH}/\eta$ ratio increases creating a wide enough range of scales such that $\eta \ll D$ and $\eta_{KH} \ll \mathcal{L}$. On the other hand, for increasing $We_\lambda$, the ratio decreases as the surface tension weakens, departing from the inertial range. In the viscous-dominated regime in Fig.~\ref{fig:regimes_interface} (b), $\eta_{KV} \ll \eta$, for increasing $Re_\lambda$, the $\eta_{KV}/\eta$ ratio increases, weakening the effect of viscous stresses on the deformation of the interface. However, for increasing $Ca_\lambda$, the ratio decreases as the surface tension weakens, strengthening the effects of viscous stresses. In an attempt to unify the evaluation of interfacial regimes in turbulence, we present the new non-dimensional number, $Ris$, which represents the ratio of interface scales in turbulence. In the Fig.~\ref{fig:regimes_interface} (c), the dotted line represents the demarcation of $Ris = 1$ such that $\eta_{KH} \sim \eta_{KV}$. For higher $Re_\lambda$, where $Ris\gg1$, the critical interface scale is in the inertial range. For higher $We_\lambda$, where $Ris\ll1$, the critical interface scale is in the viscous dissipative range.
%yielding in the limiting scale in the viscous dissipative range.

\ifstrictfloat
\begin{figure}[H]
\else
\begin{figure}%[h]
\fi
    \centering
    \begin{subfigure}{0.32\linewidth}
    \includegraphics[width=\linewidth]{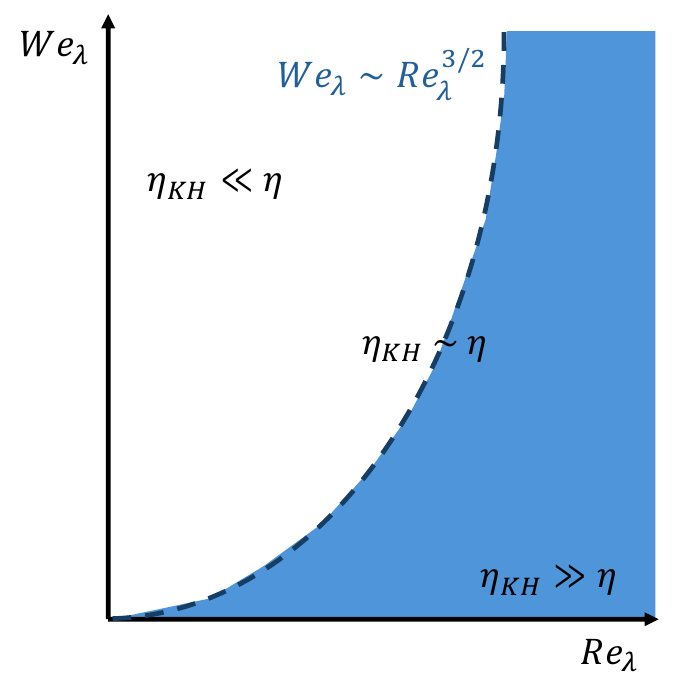}
    \caption{}
    \end{subfigure}
    \begin{subfigure}{0.32\linewidth}
    \includegraphics[width=\linewidth]{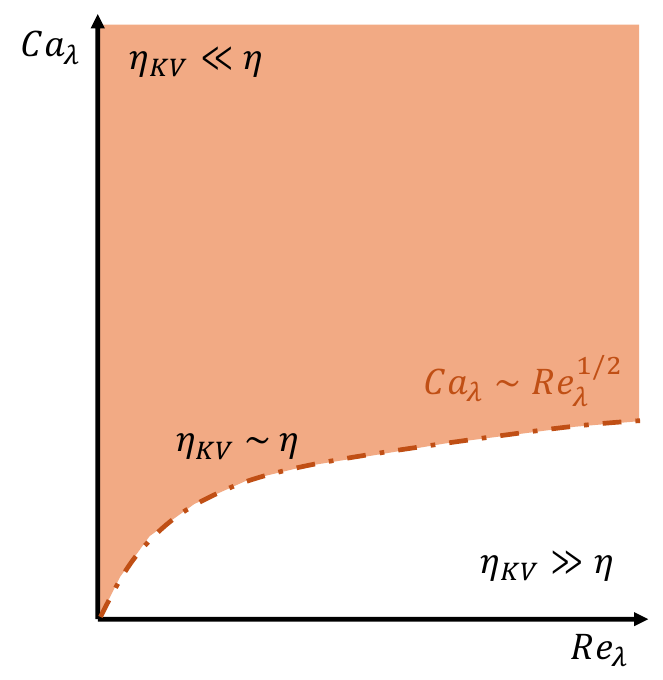}
    \caption{}
    \end{subfigure}
    \begin{subfigure}{0.32\linewidth}
    \includegraphics[width=\linewidth]{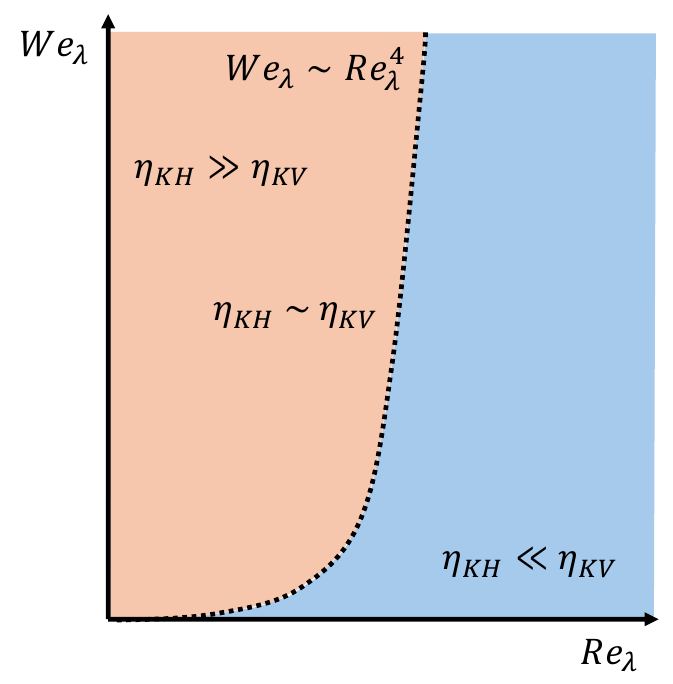}
    \caption{}
    \end{subfigure}
    \caption{Regime demarcations of interface scales in turbulence. (a) Inertia-dominated, (b) viscous-dominated, (c) unified regimes. Regimes of interest are highlighted in the two shades of blue, defining an interface in the inertial range, $\eta \ll \eta_{KH} \ll \mathcal{L}$, and, in the two shades of orange, defining a sub-Kolmogorov scale interface, $\eta_{KH} \ll \eta$. Reference scalings for order of unity ratios in dashed blue line for $\eta_{KH}/\eta \sim 1$, in dash-dot orange line for $\eta_{KV}/\eta \sim 1$, and in dotted black line for $\eta_{KH}/\eta_{KV} \sim 1$.}
    \label{fig:regimes_interface}
\end{figure}%\todo[]{use different coloring in (c), it is confusing. You say light blue is inertial range, is this true in (a)? Similarly for orange in (b)?}

\section{Resolution requirements for direct simulations of two-phase flows}
\label{sec:reqs}

% - derive grid requirements in terms of N;
Grid-point and time-step requirements are derived here in the context of two-phase flow simulations. First, in the Subsection~\ref{subsec:grid_req}, single-phase grid-point requirements are extended to two-phase, considering the constraints imposed by the presence of the dispersed phase in a turbulent flow field. Second, time-step requirements are evaluated in the Subsection~\ref{subsec:time_req} also taking into account the stability bounds imposed by the interface-tracking/capturing method.

\subsection{Grid-point requirement}
\label{subsec:grid_req}

First, a generic derivation of the number of grid points, $N_{\rm total, ref}$, required for direct numerical simulations is presented for a reference length scale, $\eta_{\rm ref}$. Second, the corresponding scalings are computed for single-phase/carrier turbulence, two-phase inertia-dominated, and viscous-dominated regimes.

Direct numerical simulations of turbulent flows require the grid spacing to be on the order of the reference length scale, $\eta_{\rm ref}$. %~\cite{yeung:2018}.  
Oftentimes, this requirement is expressed in terms of the resolution $k_{\max} \eta_{\rm ref}$, where $k_{\max}$ is the maximum wavenumber resolved by the grid. The total number of grid points, $N_{\rm total, ref}$, is as follows
\begin{equation}
    \label{eq:total_grid_1}
    N_{\rm total, ref} = N_{x, \rm ref} N_{y, \rm ref} N_{z, \rm ref},
\end{equation}
where $N_{x_i, \rm ref}$ is the total number of grid points in the $x_i$\textsuperscript{th} direction imposed by the reference scale. For HIT, the requirements in all directions are the same, given the translation and rotation/reflection invariance of the flow field, hence, the total number of un-nested grid points simplify to
\begin{equation}
    \label{eq:total_grid_2} 
    N_{\rm total, ref} = N_{x, \rm ref}^3.
\end{equation}
It is then left to determine $N_{x, \rm ref}$ as a function of the governing dimensionless numbers. By definition,
\begin{equation}
    \label{eq:nx_def} 
    N_x = \frac{L_x}{\Delta},
\end{equation}
where $L_x$ is the domain size and $\Delta$ is the grid spacing. For a periodic dimension, the grid spacing can be expressed in terms of the resolution $k_{\max} \eta_{\rm ref}$
\begin{equation}
    \label{eq:delta_gen}
    \Delta_{\rm ref} = \frac{\pi}{k_{\max} \eta_{\rm ref}} \eta_{\rm ref}.
\end{equation}
Substitute Eq.~\eqref{eq:delta_gen} in~\eqref{eq:nx_def} to obtain the number of grid points imposed by the reference length scales in the $x$ direction
\begin{equation}
    \label{eq:nx_gen}
    N_{x{\rm , ref}} = L_x \frac{k_{\max} \eta_{\rm ref}}{\pi} \frac{1}{\eta_{\rm ref}} = \frac{L_x}{\mathcal{L}_x} \frac{k_{\max} \eta_{\rm ref}}{\pi} \frac{\mathcal{L}_x}{\eta_{\rm ref}}.
\end{equation}
Therefore, given the reference length scales for single-phase/carrier turbulence, $\eta$, two-phase inertia-dominated, $\eta_{KH}$, and viscous-dominated, $\eta_{KV}$, regimes, the total number of grid points imposed by each scale can be computed using the scalings presented in Table~\ref{tab:summary_multi_turb_scales}. The total number of grid points imposed by the viscous length scales in turbulence is
\begin{eqnarray}
    \label{eq:total_grid_visc_rel} 
    N_{\rm total, turb} = N_{x{\rm, turb}}^3 &=& \left (\frac{L_x}{\mathcal{L}_x} \right )^3 \left ( \frac{k_{\max} \eta}{\pi} \right )^3 Re_\mathcal{L}^{9/4}, \\
    \label{eq:total_grid_visc_relambda}
    &=& \left (\frac{3}{20} \right )^{9/4} \left (\frac{L_x}{\mathcal{L}_x} \right)^3 \left ( \frac{k_{\max} \eta}{\pi} \right )^3 Re_\lambda^{9/2},
\end{eqnarray}
where $\mathcal{L}/L$ is the ratio of the large to domain length scales.
% can be taken as $0.2$~\cite{rosales:2005,palmore:2018}.
In single-phase turbulent flows the grid spacing is required to be on the order of the Kolmogorov length scale, $\eta$, thus, $k_{\max} \eta$ is oftentimes taken to be at least 1.5~\cite{yeung:2018}.

Moreover, the total number of grid points imposed by the interface length scales, in the inertia-dominated regime, is
\begin{eqnarray}
    N_{\rm total, surf} &=& N_{x{\rm, surf}}^3, \nonumber\\
    \label{eq:total_grid_surf_wel} 
    &=& \left (\frac{3}{2} \right )^{9/5} C_2^{6/5}\left (\frac{L_x}{\mathcal{L}_x} \right )^3 \left ( \frac{k_{\max} \eta_{KH}}{\pi} \right )^3 \left (\frac{We_\mathcal{L}}{We_t^c} \right )^{9/5}, \\
    \label{eq:total_grid_surf_welambda_relambda}
    &=& \left (\frac{27}{800} \right )^{9/10} C_2^{6/5} \left (\frac{L_x}{\mathcal{L}_x} \right)^3 \left ( \frac{k_{\max} \eta_{KH}}{\pi} \right )^3 \left (\frac{We_\lambda Re_\lambda}{We_t^c} \right )^{9/5},
\end{eqnarray}
where $C_2$ is the inertial range constant, $C_2 = u_r^2/(\epsilon r)^{2/3}$ for $\eta \ll r \ll \mathcal{L}$. Herein, the parameter $k_{\max} \eta_{KH}$ is introduced, which will also be the object of the verification study.

Lastly, the total number of grid points imposed by the interface length scales, in the viscous-dominated regime, is
\begin{eqnarray}
    N_{\rm total, cap} &=& N_{x{\rm, cap}}^3, \nonumber\\
    \label{eq:total_grid_surf_capl} 
    &=& \left (\frac{L_x}{\mathcal{L}_x} \right )^3 \left ( \frac{k_{\max} \eta_{KV}}{\pi} \right )^3 \left (\frac{Ca_\mathcal{L}}{Ca_t^c} \right )^{3}, \\
    \label{eq:total_grid_surf_calambda_relambda}
    &=& \left (\frac{3}{200} \right )^{3/2} \left (\frac{L_x}{\mathcal{L}_x} \right)^3 \left ( \frac{k_{\max} \eta_{KV}}{\pi} \right )^3 \left (\frac{Ca_\lambda Re_\lambda}{Ca_t^c} \right )^{3}.
\end{eqnarray}

The conservative approach is to take the total number of grid points as
\begin{equation}
    \label{eq:ntotal_visc_surf}
    N_{\rm total} = \max \left (N_{\rm total, turb} ; N_{\rm total, surf}  ; N_{\rm total, cap} \right )
\end{equation}

%\todo[inline]{can probably couple both interface resolutions via the Ris number}

\subsection{Time-step requirement}
\label{subsec:time_req}

Estimation of time-step requirements follows the same approach as proposed by Yang and Griffin (2021)~\cite{yang:2021a}. The total number of time steps is directly related to the cost of the simulation, since it dictates the number of iterations that the governing equations are required to be integrated in time, determining total number of FLOPS, thus cost. Within the context of stationary turbulent flows,
\begin{equation}
\label{eq:time_step_req}
\begin{aligned}
\text{Time-step } & \text{requirement} \\
& = \frac{\text{Physical time needed for converged statistics}}{\text{Time-step size}}.
\end{aligned}
\end{equation}
First, the time-step size is estimated based on the numerical stability criteria. For genericity, consider that all terms in the governing equations are treated explicitly, 
therefore, 
%for the phase field model applied to the Navier-Stokes equations (described in the Subsection~\ref{subsec:setup}), 
there will be 3 criteria: one for the convective, one for the diffusive, and one for the interface-tracking/capturing terms. In all cases, stability requires the Courant-Friedrichs-Lewy (CFL) number to be smaller than some $\mathcal{O}(1)$ number. For DNS of an incompressible two-phase flow solver with \textit{e.g.}, a phase-field model,
% in three dimensions, 
the 3 CFL criteria are
\begin{eqnarray}
    \label{eq:cfl}
    {\rm CFL_{conv}} &=& \frac{|u_{\max}| dt}{\Delta}, \nonumber\\
    {\rm CFL_{visc}} &=& \frac{\nu dt}{\Delta^2}, \\
    {\rm CFL_{surf}} &=& \frac{\Gamma \varepsilon dt}{\Delta^2}, \nonumber
\end{eqnarray}
where $\nu$ is the kinematic viscosity, $\Gamma=\Gamma^o |u_{\max}|$ is the interface velocity, and $\varepsilon=\varepsilon^o \Delta$ is the interface thickness.
In most interface-tracking/capturing methods, the ${\rm CFL_{surf}}$ can be expressed as ${\rm CFL_{surf}} = \alpha{\rm CFL_{conv}}$. For example, in the accurate diffuse interface (ACDI)~\cite{jain:2022a} method, ${\rm CFL_{surf}} = \Gamma^o \varepsilon^o {\rm CFL_{conv}}$. Therefore, the convective and interface time-step requirements are related as, \textit{i.e.}, ${dt_{\rm surf}} = {dt_{\rm conv}} / \alpha$. The interface stability would be more restrictive if $\alpha > 1$. Consider the general case, if we assume ${\rm CFL} \leq 1$, then, from Eq.~\eqref{eq:cfl},
\begin{equation}
    \label{eq:dts_1}
    dt \leq \min \left (\frac{1}{\alpha} \frac{\Delta}{|u_{\max}|}; \frac{\Delta^2}{\nu}\right ).
\end{equation}
%
%The convective time-step stability bound can be computed with $\alpha = 1$.
%Stability bounds for the accurate conservative diffuse interface (ACDI) phase field model~\cite{jain:2022a} require $\Gamma^o=1$ and $\epsilon^o>0.5$. 
%For $dt_{\rm conv} < dt_{\rm surf}$, $\epsilon^o < 1/6$, contradicting the stability criteria. Therefore, Eq.~\ref{eq:dts_1} simplifies to
%
%\begin{equation}
%    \label{eq:dts_2}
%    dt \leq \min \left (\frac{\Delta^2}{6\nu}; \frac{\Delta}{6 |u_{\max}| \varepsilon^o}\right ).
%\end{equation}
%

The remainder of the derivation of the total number of time-step, $N_t$, is rather long, since one has to consider all possible combinations of the three limiting grid resolutions with at least two time-step size restrictions, the convective/interface, and the viscous ones. Therefore, only scalings for $N_t$ are presented here (for the complete derivation of the minimum time-step sizes, as well as the total number of time steps to be integrated, see Appendix~\ref{apx:time_step_reqs_complete}).

%\todo[inline]{add summary table}

Total physical time to attain statistical convergence is estimated based on the eddy-turnover time~\cite{pope:2000,bassenne:2016}, $\tau_e$,
\begin{equation}
    \label{eq:tau_e}
    \tau_e = \frac{2k}{3\epsilon} = \frac{\mathcal{L}}{u'} \frac{u'^2}{\nu} \frac{\nu}{u'^2} = \sqrt{\frac{2}{3}} \frac{Re_\mathcal{L} }{u'^2/\nu}.
\end{equation}
Consider $C_t$ to be the number of $\tau_e$ to be integrated such that the statistics are converged, hence, according to Eq.~\eqref{eq:time_step_req}, the total number of time steps,$N_{t,\rm ref}$, based on a reference $dt_{\rm ref}$, is
\begin{equation}
    \label{eq:nt}
    N_{t, \rm ref} = \frac{C_t \tau_e}{dt_{\rm ref}} = \sqrt{\frac{2}{3}} \frac{C_t Re_\mathcal{L}}{u'^2 dt_{\rm ref}/\nu}.
\end{equation}
%
%\todo[inline]{add explanations}
Therefore, a lower bound for $N_t$ can be constructed based on all possible combinations of the minimum grid size and the minimum time-step size. Given all time step sizes derived in the Appendix~\ref{apx:time_step_reqs_complete}, substitute in Eq.~\eqref{eq:nt} to obtain the scalings (for the explicit expressions refer to the Appendix~\ref{apx:time_step_reqs_complete}) for the total number of time steps as functions of dimensionless number $Re$, $We$, $Ca$, and the resolutions for each regime $k_{\max} \eta$, $k_{\max} \eta_{KH}$, $k_{\max} \eta_{KV}$. All results are summarized in the Table~\ref{tab:summary_multi_turb_nt}.
\ifstrictfloat
\begin{table}[H]%% placement specifier
\else
\begin{table}%[H]%% placement specifier
\fi
\centering%% For centre alignment of tabular.
\caption{Power laws for total number of time step, $N_t$, in turbulence, inertia-dominated, and viscous-dominated regimes bounded by convective/interface and viscous stability criteria as functions of $Re$, $We$, $Ca$, and $\Delta_{\rm turb}$ ($k_{\max} \eta$), $\Delta_{\rm surf}$ ($k_{\max} \eta_{KH}$), and $\Delta_{\rm cap}$ ($k_{\max} \eta_{KV}$).}\label{tab:summary_multi_turb_nt}
\begin{tabular}{c | c | c}%% Table column specifiers
  \hline
                                            & $N_{t, \rm conv/surf}$ & $N_{t, \rm visc}$                         \\ \hline
   $\Delta_{\rm turb} (k_{\max} \eta)$      & $Re_\mathcal{L}^{3/4}$ & $Re_{\mathcal{L}}^{1/2}$                  \\
   $\Delta_{\rm surf} (k_{\max} \eta_{KH})$ & $We_\mathcal{L}^{3/5}$ & $We_\mathcal{L}^{6/5}Re_\mathcal{L}^{-1}$ \\
   $\Delta_{\rm cap} (k_{\max} \eta_{KV})$  & $Ca_\mathcal{L}$       & $Ca_\mathcal{L}^{2}Re_\mathcal{L}^{-1}$   \\ \hline
\end{tabular}
\end{table}
Overall for increased $Re$, $We$ and $Ca$, the total number of time steps required for converged statistics increases, as expected. Although $C_t$ is likely not to change significantly with the governing parameters, $dt$ is required to be reduced because the range of scales increases with the decrease in the smallest length and their corresponding time scales. Therefore, it becomes computationally more intensive both in terms of memory and wall-clock time as $Re$, $We$, and $Ca$ are increased. Interestingly, for the total number of time steps imposed by the viscous CFL and the interface scales, $N_t$ is reduced with $Re$. This is due to the fact that both $\eta_{KH}/\eta$ and $\eta_{KV}/\eta$ are increasing functions of $Re$; therefore, as $Re$ increases, it is more likely that the carrier turbulence scale will become more restrictive than the interface scales, thereby reducing $N_{t, \rm visc, surf}$ and $N_{t, \rm visc, cap}$, and increasing $N_{t, \rm turb}$, eventually becoming the most demanding.

\section{Verification using numerical simulations}
\label{sec:verify}

% - connect with bubble rising requirements, now Hinze acts like the most relevant length scale (Batistella et al, 2020, derived in quiescent) - also try to find more papers;
% - in the bubble rising, they said we need 20 points, now similar still to have 20 points but on the Hinze scale instead of the rising bubble;
% - table listing parameters and range

% Writing starts here:
% 1) governing equations
\subsection{Governing equations}
\label{subsec:gov_eq}

%\todo[inline]{review setup sections, copied them from ILASS}

This work focuses on incompressible two-phase turbulent flows with linear forcing to maintain the stationary state of the turbulence. The accurate conservative diffuse-interface/phase-field (ACDI) model~\cite{jain:2022a} is chosen as the interface-capturing method, written as
\begin{align}
\label{eq:vol_transp}
\frac{\partial \phi_m}{\partial t} + \frac{\partial \phi_m u_j}{\partial x_j} = \frac{\partial a_{m,j}}{\partial x_j} ,
\end{align}

\noindent where $\phi_m$ is the volume fraction of phase $m=1,2$, $u_i$ is the velocity component, and $a_{m,i}$ the interface-sharpening flux of the respective phase $m$ in the $i$\textsuperscript{th} direction. The interface-sharpening flux maintains the interface equilibrium $\tanh$ profile. The definition of $a_{m,i}$ is
\begin{align}
\label{eq:int_sharp}
a_{m,i} = \Gamma \Bigg [ \varepsilon \frac{\partial \phi_m}{\partial x_j}  + \frac{1}{4} \left (1 - \tanh^2 \left (\frac{\psi}{2\varepsilon} \right ) \right ) n_j  \Bigg ], 
\end{align}

\noindent where $\Gamma$ is the interface velocity, set as $\max(\|u_i\|)$, $\varepsilon$ is the interface thickness, $\psi$ is an auxiliary sign-distance function defined as $\psi = \ln [(\phi + 10^{-100})/(1-\phi + 10^{-100})]$, which varies more smoothly than $\phi$ thus more accurately computing interface normals, $n_i$, and interface curvature, $\kappa$. The interface normal is defined as $\displaystyle n_i = \frac{\partial \psi}{\partial x_i} \Bigg /  \sqrt{\frac{\partial \psi}{\partial x_k}\frac{\partial \psi}{\partial x_k}}$.

The divergence-free condition and the consistent momentum transport equations are 
%respectively, presented:
%
\begin{align}
\label{eq:gov_eqs}
& \frac{\partial u_i}{\partial x_i}  = 0, \\ 
& \begin{aligned}
\frac{\partial \rho u_i}{\partial t} + \frac{\partial \rho u_i u_j}{\partial x_j} = & -\frac{\partial p}{\partial x_i} + \frac{\partial \tau_{ij}}{\partial x_j} + \frac{\partial u_i f_j}{\partial x_j} 
 + \sigma \kappa \frac{\partial \phi}{\partial x_i} + F_i .
\end{aligned}
\end{align}
%Due to the incompressibility of the two fluids, the mixture density $\rho$ requires the divergence of the velocity field to be zero. 
%Conservation of mass implies the divergence-free condition on the velocity field. 
In the momentum transport equation, $\rho$ is the mixture density, $p$ is the pressure, $\tau_{ij}$ are the viscous stresses defined as $\tau_{ij} = \mu (\partial u_i/\partial x_j + \partial u_j/\partial x_i)$ for the corresponding mixture dynamic viscosity, $\mu$. The consistent formulation of the phase-field model requires a momentum regularization term at the interface, $u_i f_j$, where $f_i = \rho_1 a_{1,i} + \rho_2 a_{2,i} = (\rho_1 - \rho_2) a_{1,i}$. 
The surface tension force is modeled using the continuum
%Herein the non-conservative formulation of the continuous 
surface force approach by Brackbill \textit{et al.}~\cite{brackbill:1992}, where $\sigma$ is the surface tension and $\kappa$ the interface curvature. The latter is computed based on the gradient of the interface normal as $\kappa = - \partial n_i / \partial x_i$, where $\psi$ is used to compute the normals in the surface tension force, instead of $\phi$, which has been shown to result in reduced spurious currents \cite{jain:2022a}. The last term corresponds to linear forcing in physical space, defined as $F_i= \rho A u_i$ \cite{jain:2025}.

% 1) numerical methods
\subsection{Computational setup}
\label{subsec:setup}

A set of incompressible two-phase stationary homogeneous isotropic turbulence simulations are performed. The flow and all boundary conditions are periodic in the three dimensions. A finite-volume approach on a structured Cartesian grid is adopted for spatial discretization of the governing equations, where the velocity is staggered with respect to other variables to prevent pressure checkerboarding and to achieve discrete kinetic energy conservation. A low-dissipative second-order skew-symmetric with implicit kinetic energy preservation scheme is used, as it was shown to be robust even in the limit of infinite Reynolds number ($Re$)~\cite{jain:2022a,jain:2022b}. Time integration is performed with a fourth-order Runge-Kutta scheme. Continuity equation is enforced via a modified fractional-step method, where the pressure projects the velocity onto a divergence-free field at each sub-time step after temporal evolution of intermediate velocity field. Simulations are forced to maintain a stationary state using the approach of Jain and Elnahhas~\cite{jain:2025}, where the mixture turbulent kinetic energy is maintained constant. Simulations are performed using the in-house GPU-enabled ExaFlow3D solver~\cite{jain:2020,jain:2022b}.

A single-phase simulation is first performed for about $20\tau_e$ and then the secondary phase is introduced into the domain. With the exception of the total area evolution where the initial transient of the simulation is considered, all statistics are collected after the simulations attain steady state by the balance of breakup and coalescence, \textit{i.e.}, total area fluctuates around an average value. Two combinations of Reynolds, $Re_\lambda$, and Weber, $We_l$, numbers are considered: $Re_\lambda = 55, 87$, and $We_\mathcal{L} = 6.5, 60$. The Weber numbers are chosen such that the breakup is allowed to occur, namely above the critical Weber number, $We_t^c$, which is approximately equal to 3~\cite{qian:2006}. The Taylor-scale Reynolds number is defined by $Re_\lambda = \sqrt{20 \rho k_\infty^2 / 3 \mu \epsilon_\infty}$, which is based on the stationary turbulent kinetic energy, $k_\infty = 1$, and the energy dissipation rate, $\epsilon_\infty = (2k_\infty/3)^{3/2}/\mathcal{L} \approx 0.433$. Analogously, the large-scale Weber number is defined by $We_\mathcal{L} = 2 \rho k_\infty^{5/2} / 3 \sigma \epsilon_\infty$. 

The delimiting grid resolution value is characterized by varying the value of $k_{\max} \eta_{KH}$ whilst maintaining turbulence well resolved, \textit{i.e.}, $k_{\max} \eta > 1.5$, except for the under-resolved cases A0 and B0. The grid resolution values are chosen to verify grid convergence between $k_{\max} \eta_{KH}=60$ and $120$, which approximately translates to a requirement of about $20$ grid points over the Kolmogorov-Hinze scale. This is consistent with the current practices in two-phase flows where $20$ points is assumed to be required across the diameter of drops/bubbles to accurately capture the forces on them~\cite{dodd:2016,battistella:2020,crialesi:2022,crialesi:2023}.
%\todo{add more references}. 
Furthermore, all cases are performed with constant values for void fraction, $\langle \phi_d \rangle = 0.065$, density ratio, $\rho_d/\rho_c = 1$ (for simplicity, $\rho = 1$), and, viscosity ratio, $\mu_d/\mu_c = 1$ ($\mu$ adjusted based on $Re_\lambda$). All cases are shown in Table~\ref{tab:cases_inertia}. The parameter $\eta_{KH}/\eta$ highlights the relative ratio between the inertial-turbulent interface and single-phase scales. As this ratio decreases, the Kolmogorov-Hinze scale departs from the inertial range and viscous shear stress starts to interfere in the breakup processes. The resolution requirements may vary for viscous-dominated regimes, therefore, the ratio $\eta_{KH}/\eta$ is limited to sufficiently high values to be fully in the inertia-dominated regime. The time-step size is chosen such that the CFL number is about 0.5. The simulations are integrated for up to a minimum of $20$ eddy-turnover times ($\tau_e = 2k_\infty/3\epsilon_\infty$). The transient period spans approximately the initial $5 \tau_e$, and statistics are collected over a minimum of $15 \tau_e$.

%\todo[inline]{comment on dt, holding CFL approx 0.5 with constant time step}

%\todo[inline]{check capillary number for this section (not to include in the ASME presentation}

%\todo[inline]{add A0/E0 case when turbulence is not resolved}

\ifstrictfloat
\begin{table}[H]% add [H] placement to break table across pages
\else
\begin{table}%[H]% add [H] placement to break table across pages
\fi
\caption{\label{tab:cases_inertia}Parameters of the inertia-dominated, $\eta_{KH} > \eta$, runs in dimensionless units. $We_\mathcal{L}$, integral-scale Weber number. $\eta_{KH} / \eta$ ratio of interface to turbulence scales in the inertial-dominated regime. $k_{\max}\eta$, viscous resolution based on the Kolmogorov scale $\eta$ and the maximum wave number to be discretized in the specified number of grid points. $k_{\max}\eta_{KH}$, grid resolution based on Kolmogorov-Hinze scale $\eta_{KH}$ and the maximum wave number. All quantities are rounded to three significant figures.}
\begin{ruledtabular}
\begin{tabular}{c c c c c c c c}
Run & A0 & A1 & A2 & B0 & B1 & B2 & B3                                    \\ \hline
$Re_\lambda$            & \multicolumn{3}{c}{55}   & \multicolumn{4}{c}{87}  \\
$We_\mathcal{L}$        & \multicolumn{3}{c}{6.5}  & \multicolumn{4}{c}{60}  \\
$\eta_{KH}/\eta$        & \multicolumn{3}{c}{40.8} & \multicolumn{4}{c}{21.4}  \\
%$\langle \phi_d\rangle$ & \multicolumn{8}{c}{0.065} \\
%$\rho_d/\rho_c$         & \multicolumn{8}{c}{1}     \\
%$\mu_d/\mu_c$           & \multicolumn{8}{c}{1}     \\
$k_{\max} \eta$         & 0.38 & 1.50 & 3.00              & 0.76 & 1.51 & 3.02 & 6.04               \\
$k_{\max} \eta_{KH}$    & 15.3 & 61.3 & 122.6             & 16.2 & 32.3 & 64.6 & 129.2        
\end{tabular}
\end{ruledtabular}
\end{table}%\todo[]{It might be better to add a vertical line in the table  between A2 and B0.}

\subsection{Results}
\label{subsec:results}

% outline:
% - talk about quantities of interest - bubble distribution, total area, pdf of curvature, SMD;
% - for total area - one plot on time evolution for different resolutions and and include some regions that go to steady state. and say that here onwards we use only the state state part for the other quantities;

% Writing starts here:
% 1) Cases

% 2) Results
% a) Interfacial area
%\todo[inline]{change transient graph to inset}

%\todo[inline]{Add a line or two listing the four quantities here.}

In order to determine and verify the adequate grid resolution for high-fidelity simulations of two-phase flows, convergence of several quantities of interest (interfacial area, size distribution, SMD, and curvature distribution) are evaluated in the following subsections.
%: interfacial area in Section~\ref{subsubsec:area}, size distribution in Section~\ref{subsubsec:size_dist}, SMD in Section~\ref{subsubsec:smd}, and curvature distribution in Section~\ref{subsubsec:curv}.

\subsubsection{Interfacial area}
\label{subsubsec:area}

The first quantity of interest is the interfacial area, which is a parameter that governs mass, momentum, and energy transfer between the phases. Two regimes of interest are considered: transient and stationary. The transient evolution, although dependent on the initialization procedure, is useful to determine whether the breakup is being properly captured if its evolution is converged. Following the results from Jain and Elnahhas~\cite{jain:2025}, the velocity field is not changed from the single-phase forced turbulence state when adding the dispersed phase, since this was shown to yield in faster turnover to achieve stationary state, therefore reducing the overall simulation computational cost. The stationary regime represents a state of balanced breakup and coalescence. 

Both regimes are presented in Figure~\ref{fig:area_evolution_cases} for cases A and B from Table~\ref{tab:cases_inertia}, where plots of total interfacial area, normalized by their initial value, are labeled based on their grid resolution $k_{\max} \eta_{KH}$. For the proposed void fraction $\langle \phi_d \rangle = 0.065$, the initial area is $A_0 = 4 \pi r_0^2 = \pi^3$. %Moreover, the insets on the Figures highlight the transient period of the simulations. 
It is clear that for the under-resolved cases A0 and B0, with $k_{\max} \eta < 1.5$ and $k_{\max} \eta_{KH} = 15$, both transient and stationary regimes are not properly captured. For both sets of dimensionless parameters, initial breakup is slower since the grid resolution is not sufficient to allow surface corrugations to develop that would eventually yield in breakup. Moreover, total interfacial area at steady state is also underpredicted given that smaller and intermediate bubbles/droplets cannot be formed due to the lack of corrugations and unstable surface topologies.

For both cases, the minimum resolution of $k_{\max} \eta_{KH} = 60$ is able to capture the breakup-dominated regime, as shown by the convergence between the 60 and 120 resolutions within $t/\tau_e$ from 0 to 5. This is an initial indication that, within the range of dimensionless numbers studied and for the adopted numerical methods herein, breakup resulting in small bubbles/droplets contributing to more significant amounts of interfacial area are properly
%\todo[inline]{I think this is too strong. We can only say that breakup that results in features or small droplets that contribute significantly to total area are well captured, but not necessarily all breakup events. So make it softer.} 
captured with $k_{\max} \eta_{KH} = 60$. 

For $Re_\lambda = 55$ and $We_\mathcal{L} = 6.5$, the stationary regime can be claimed to be converged in the Figure~\ref{fig:area_evolution_cases} (a), %as to be justified by the convergence of the averaged interfacial area 
although the $k_{\max} \eta_{KH} = 120$ grid resolution demonstrates some pronounced interfacial area fluctuations at longer time scales. For the second set of parameters $Re_\lambda = 87$ and $We_\mathcal{L} = 60$, during the first stationary 5 eddy-turnover times, $\tau_e$, the total interfacial area between the resolutions $k_{\max} \eta_{KH} = 60$ and $120$ are well converged. 
A similar behavior to the previous case occurs after $t/\tau_e = 10$, where there is an increase in the total interfacial area for the finest resolution. It is likely that the increase in interfacial area to be linked with intermittency and extreme events of turbulence, which can also trigger further breakup of bubbles/droplets even at sub-Kolmogorov-Hinze scales (see, Appendix \ref{apx:intermittency_area}). In later stages, $t \approx 25 \tau_e$, the proposed resolution of $k_{\max} \eta_{KH} = 60$ also demonstrates an instance of further breakdown and an increase in the total interfacial area, likely due to intermittency. Nonetheless, if the resolution is further reduced to $k_{\max} \eta_{KH} = 30$, then the grid resolution is no longer able to capture these events.
Even though turbulence resolution of $k_{\max} \eta = 1.5$ has been used as reference given its convergence of low-order moments, in order to capture intermittency and extreme events it may not be sufficient~\cite{yeung:2018}.%\todo{intermittency is more of a single/carrier phase phenomenon which requires a certain viscous resolution, how can we identify the actual interface resolution while maintaining the same viscous resolution?}

\ifstrictfloat
\begin{figure}[H]
\else
\begin{figure}%[H]
\fi
    \centering
    \begin{subfigure}{0.45\linewidth}
        \centering
        \includegraphics[width=\linewidth]{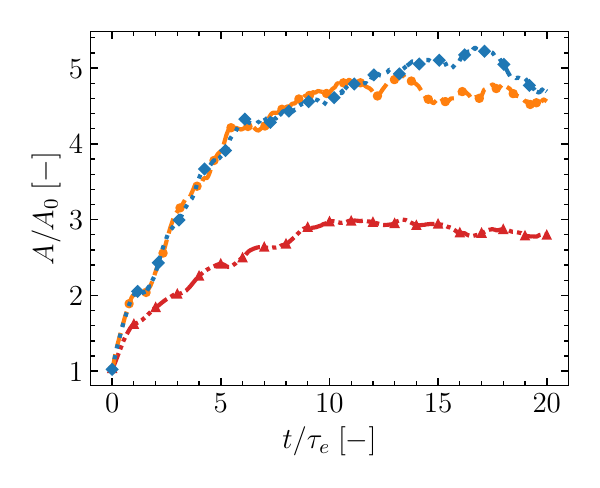}
        \caption{}
    \end{subfigure}
    \begin{subfigure}{0.45\linewidth}
        \centering
        \includegraphics[width=\linewidth]{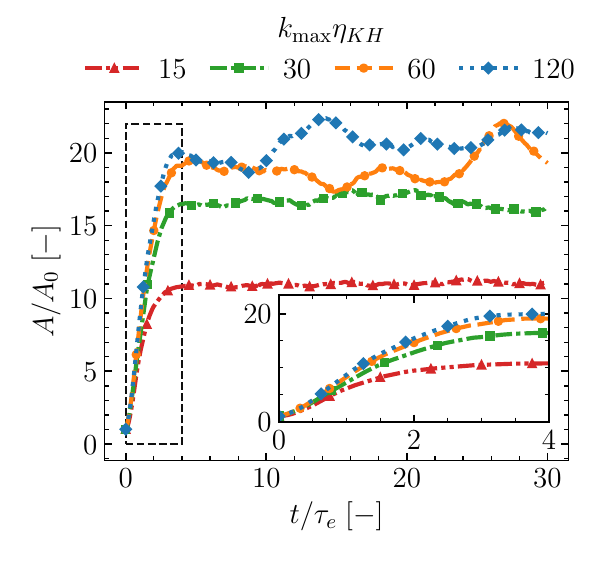}
        \caption{}
    \end{subfigure}
    \caption{Interfacial area evolution as a function of grid resolution. (a) Cases A, $Re_\lambda = 55$, $We_\mathcal{L} = 6.5$, (b) cases B, $Re_\lambda = 87$, $We_\mathcal{L} = 60$. Transient state is highlighted for (b) cases B on the inset for $t/\tau_e \in [0, 4]$.}
    \label{fig:area_evolution_cases}
\end{figure}

Figure~\ref{fig:comp_avg_area} synthesizes the average values of the interfacial area in the stationary state, namely after $t = 5 \tau_e$. 
For both cases, the under-resolved simulations significantly underpredict the total interfacial area at steady state, approximately $50\%$ of the finest resolution. 
%Their average interfacial area, which directly relates to the average surface energy, do not provide a physical representations of the proposed pair of dimensionless numbers and would yield in severe mispredictions of exchange phenomena through the interface. 
For the cases A with lower $Re_\lambda$ and $We_\mathcal{L}$, steady state is reasonably converged since the average interfacial area for $k_{\max} \eta_{KH} = 60$ is within one standard deviation of the average value from $k_{\max} \eta_{KH} = 120$. For the cases B, it is clear that resolutions of $k_{\max} \eta_{KH} \leq 30$ are not sufficient. For instance, the grid resolution of $30$ is approximately within 3 to 4 standard deviations, while the resolution of $60$ is within 1.5 standard deviations. 
As seen in Figure~\ref{fig:area_evolution_cases} (b), during the interval $t/\tau_e \in [10, 15]$, the total interfacial area for the finest resolution is higher than the other two resolved due to intermittency, hence yielding a higher average value. After $t/\tau_e=25$, the proposed resolution of $k_{\max} \eta_{KH} = 60$ also presents an increase in the total interfacial area. Therefore, it is expected that, for very long-time averaging, convergence between both the steady-state interfacial area values for the finer resolutions should improve.
%both resolutions should be converged with respect to the average interfacial area at steady state. 

\ifstrictfloat
\begin{figure}[H]
\else
\begin{figure}%[H]
\fi
    \centering
    \includegraphics[width=0.65\linewidth]{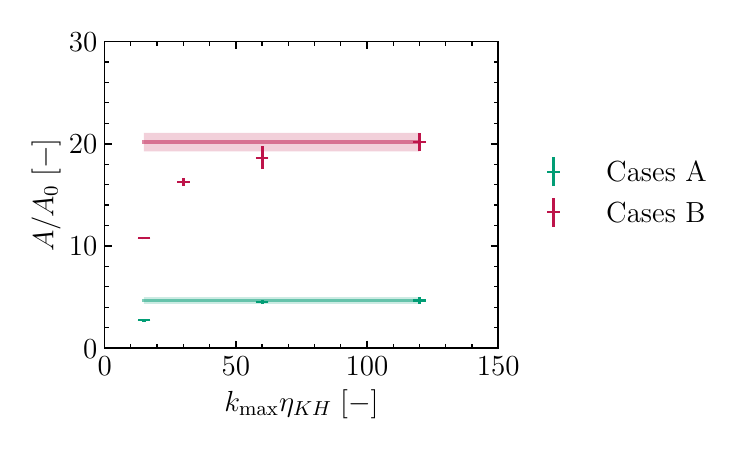}
    \caption{Average interfacial area at steady state comparison for two pairs of dimensionless number as a function of grid resolution. One standard deviation band based on finest resolution added for reference.}
    \label{fig:comp_avg_area}
\end{figure}

Visualization of the three-dimensional instantaneous flow field at $t/\tau_e=10$
is presented in Figure~\ref{fig:visualizations_3d_tau_e_10} (this specific time is chosen since the total interfacial area between $k_{\max} \eta_{KH} = 60$ and $120$ at this state are approximately equal, as can be seen in Figure~\ref{fig:area_evolution_cases} (b)). For the two sets of $Re_\lambda$ and $We_\mathcal{L}$, it is clear that the under-resolved resolutions fail to capture both turbulence and interface structures, resulting in a significant underprediction of the interfacial area. For the cases A, both grid resolutions of $60$ and $120$ seem to produce qualitatively similar structures and in approximately similar amounts. Besides the distinction between the resolved and under-resolved resolutions, the differences between other grid resolutions are less clear for cases B due to the smaller $\eta_{KH}/\eta$ ratio and higher $Re_\lambda$ and $We_\mathcal{L}$, \textit{i.e.}, the smaller bubbles/droplets are much smaller than the ones from the cases A. 
%Nevertheless, one may focus on the interface resolution $k_{\max} \eta_{KH} = 30$ compared to $k_{\max} \eta_{KH} = 60$ and $120$ for the cases B near the edges of the periodic domain, there are much less very small bubbles/droplets for the former compared with the latter two, providing additional indication that the proposed resolution is qualitatively sufficient to capture even sub-Kolmogovo-Hinze scales. 
%A further investigation on the effects of intermittency in interfacial area and the required resolution to capture these extreme events is discussed in the Appendix~\ref{apx:intermittency_area}.

\ifstrictfloat
\begin{figure}[H]
\else
\begin{figure}%[H]
\fi
    \centering
    \begin{minipage}{\textwidth}
        \centering
        \begin{tikzpicture}
            % Row of images inside a node matrix
            \matrix[column sep=2pt] (m) {
                \node[inner sep=0] (A0) {\includegraphics[width=0.24\textwidth]{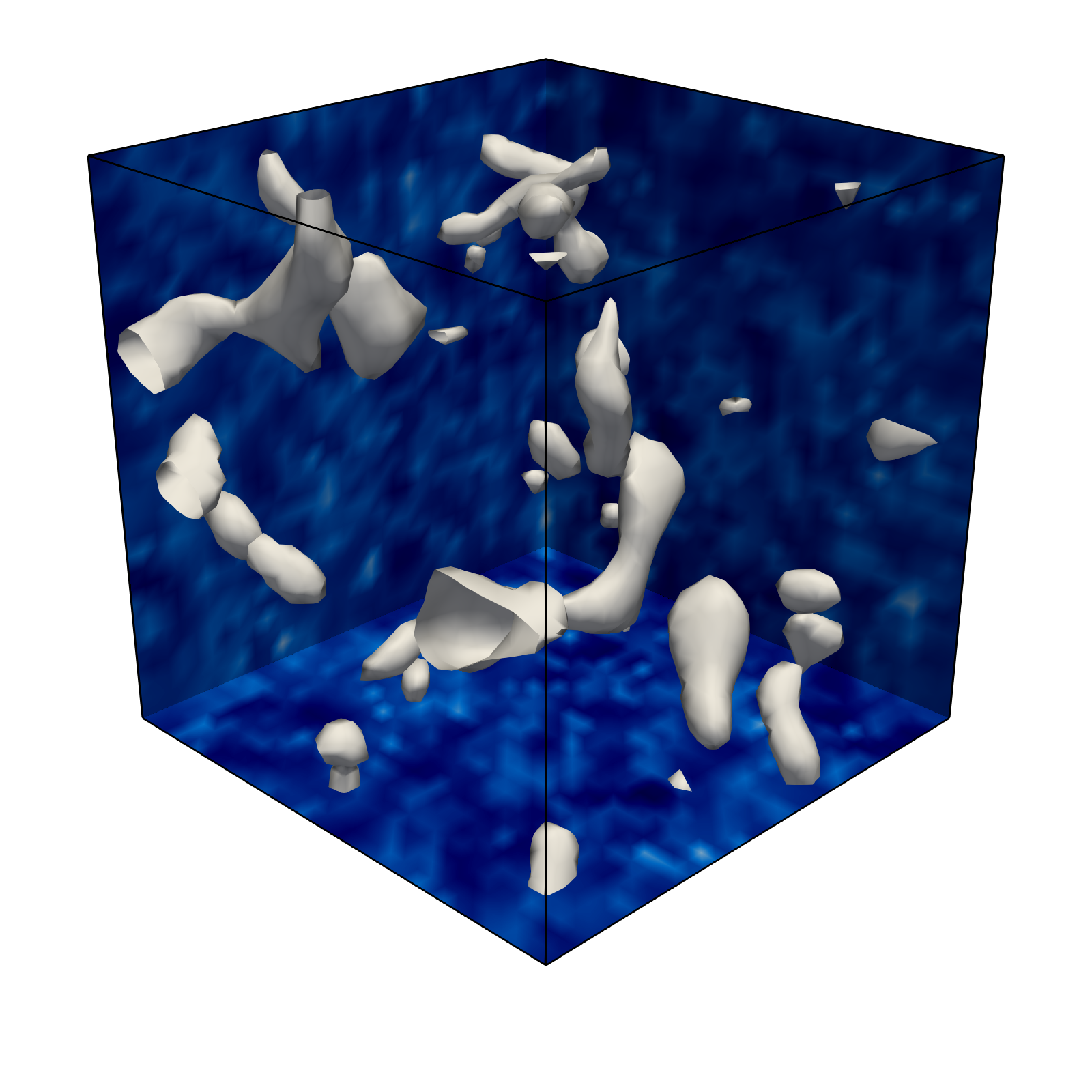}}; &
                \node[inner sep=0] (A1) {\includegraphics[width=0.24\textwidth]{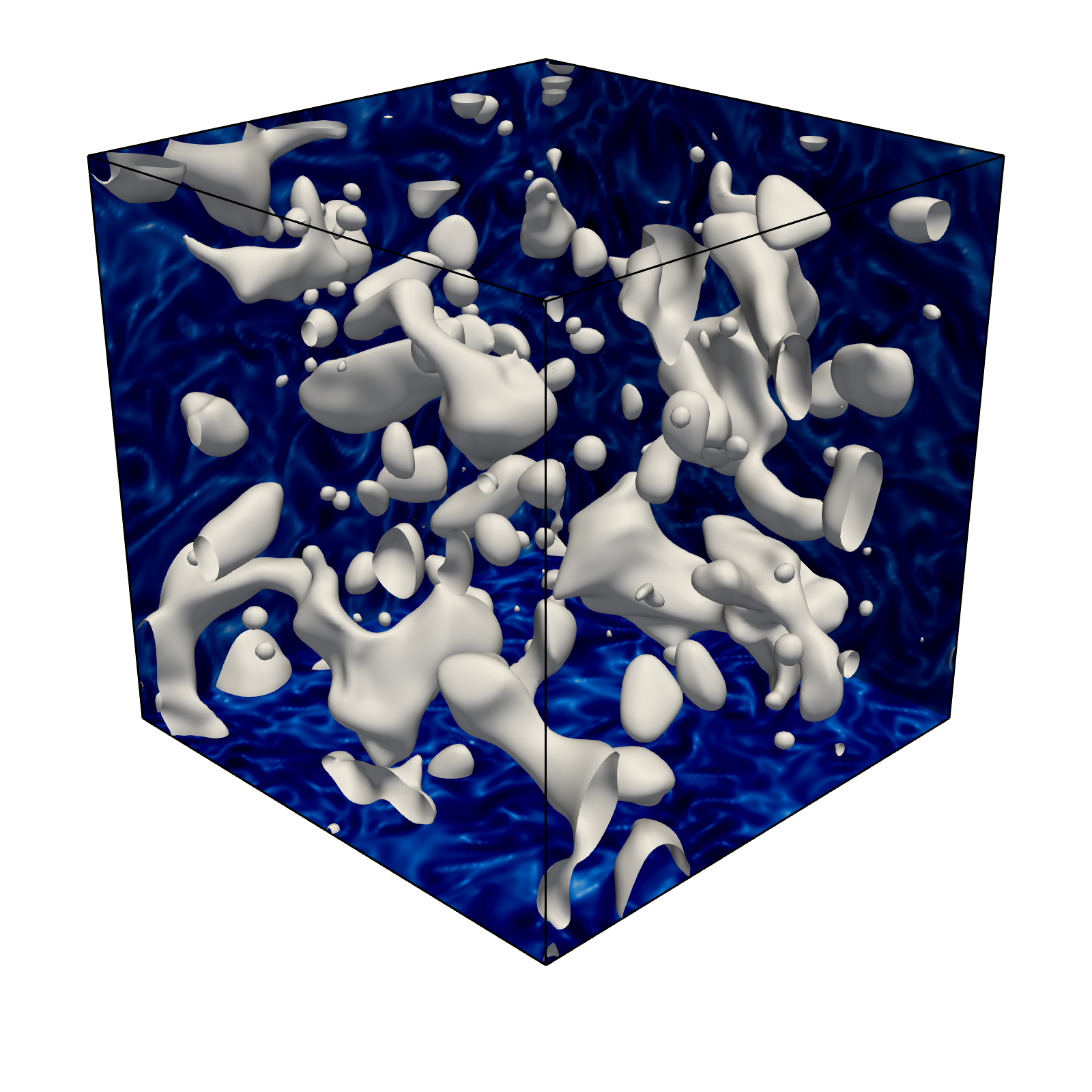}}; &
                \node[inner sep=0] (A2) {\includegraphics[width=0.24\textwidth]{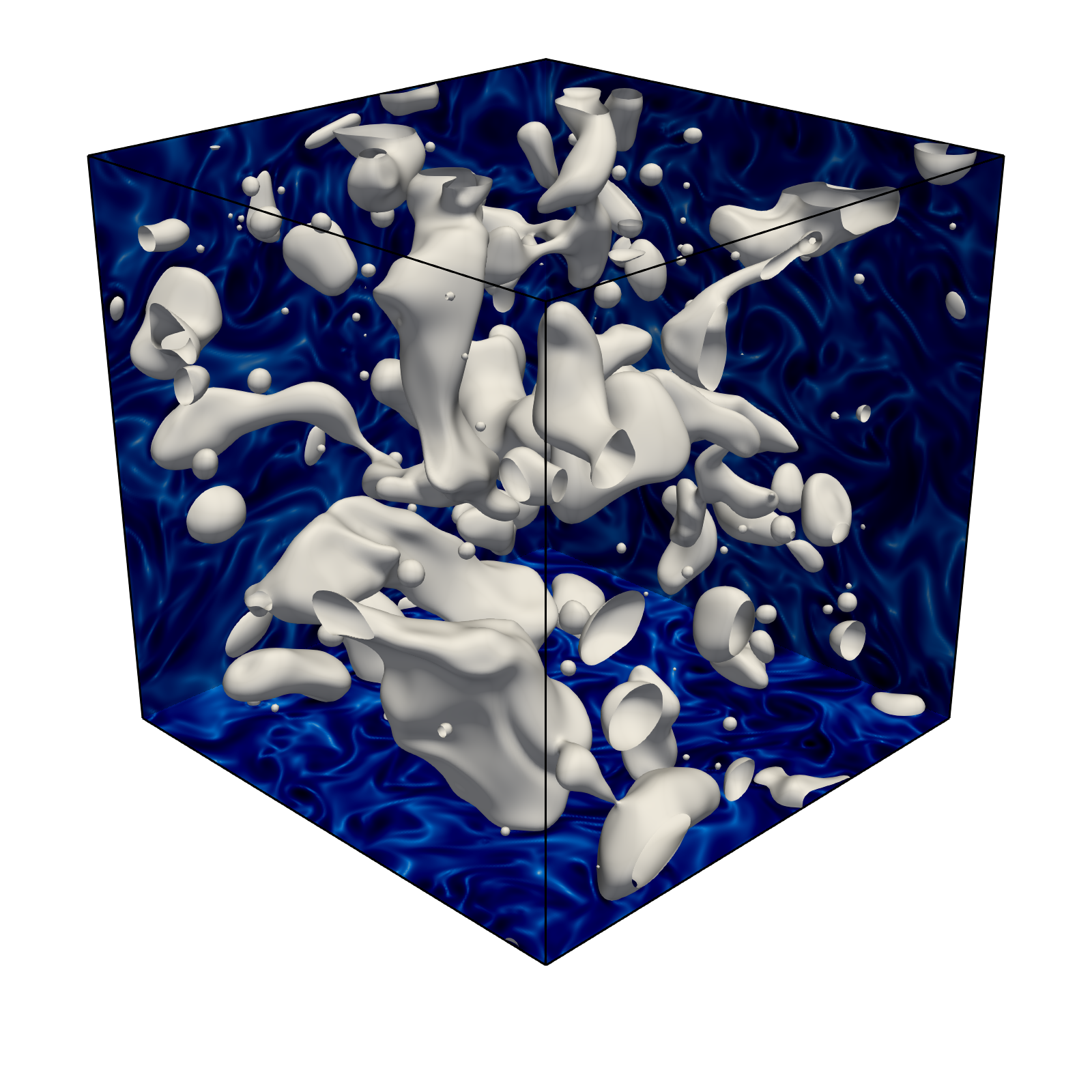}}; \\
            };
            % One rectangle around E2+E3
            \draw[blue, rounded corners, very thick] (A1.south west) rectangle (A2.north east);
        \end{tikzpicture}

        % Captions under each figure
        \parbox{0.24\textwidth}{\centering $k_{\max}\eta_{KH} = 15$}
        \parbox{0.24\textwidth}{\centering $k_{\max}\eta_{KH} = 60$}
        \parbox{0.24\textwidth}{\centering $k_{\max}\eta_{KH} = 120$}
    \end{minipage}
    
    \vspace{1em}

    \begin{minipage}{\textwidth}
        \centering
        \begin{tikzpicture}
            % Row of images inside a node matrix
            \matrix[column sep=2pt] (m) {
                \node[inner sep=0] (E0) {\includegraphics[width=0.24\textwidth]{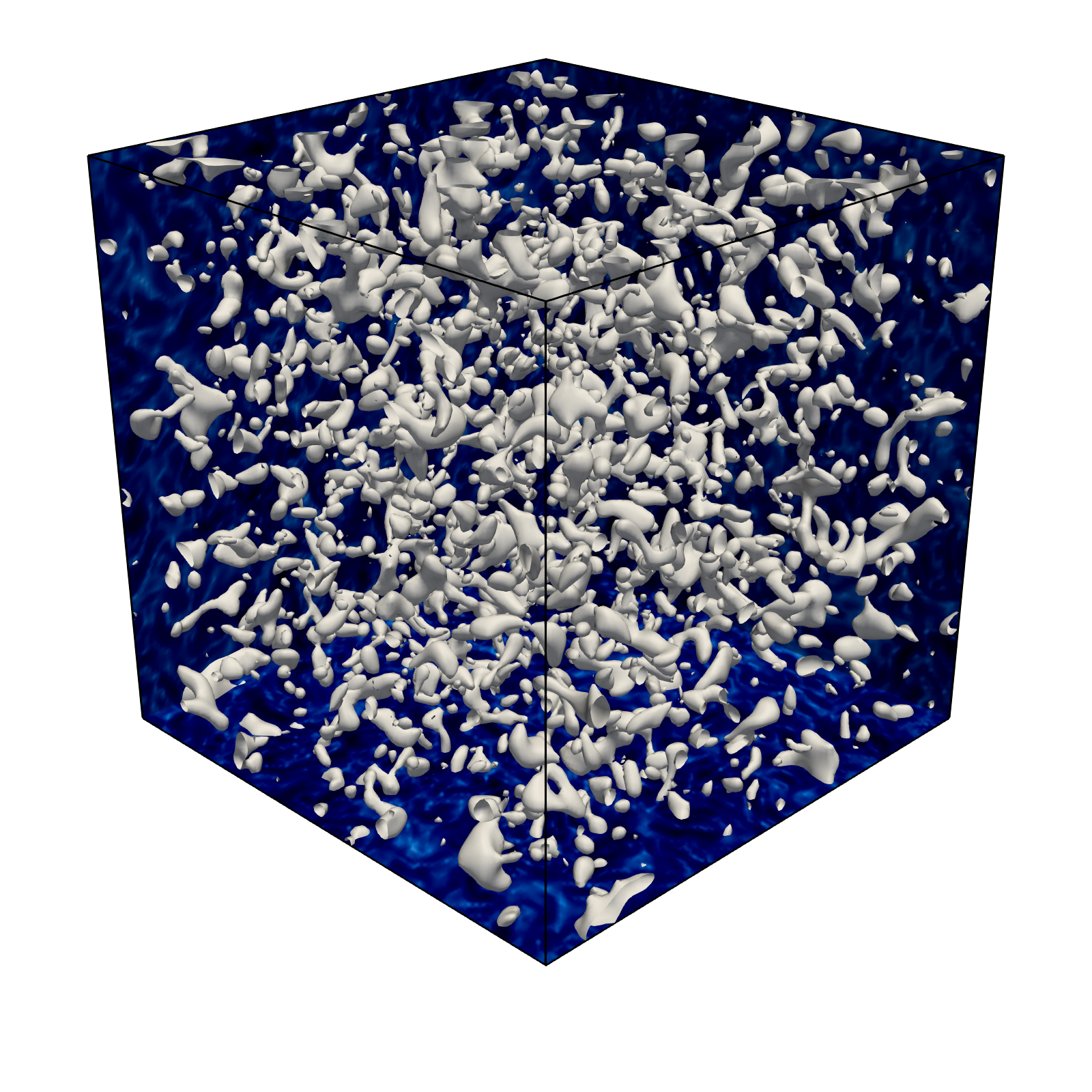}}; &
                \node[inner sep=0] (E1) {\includegraphics[width=0.24\textwidth]{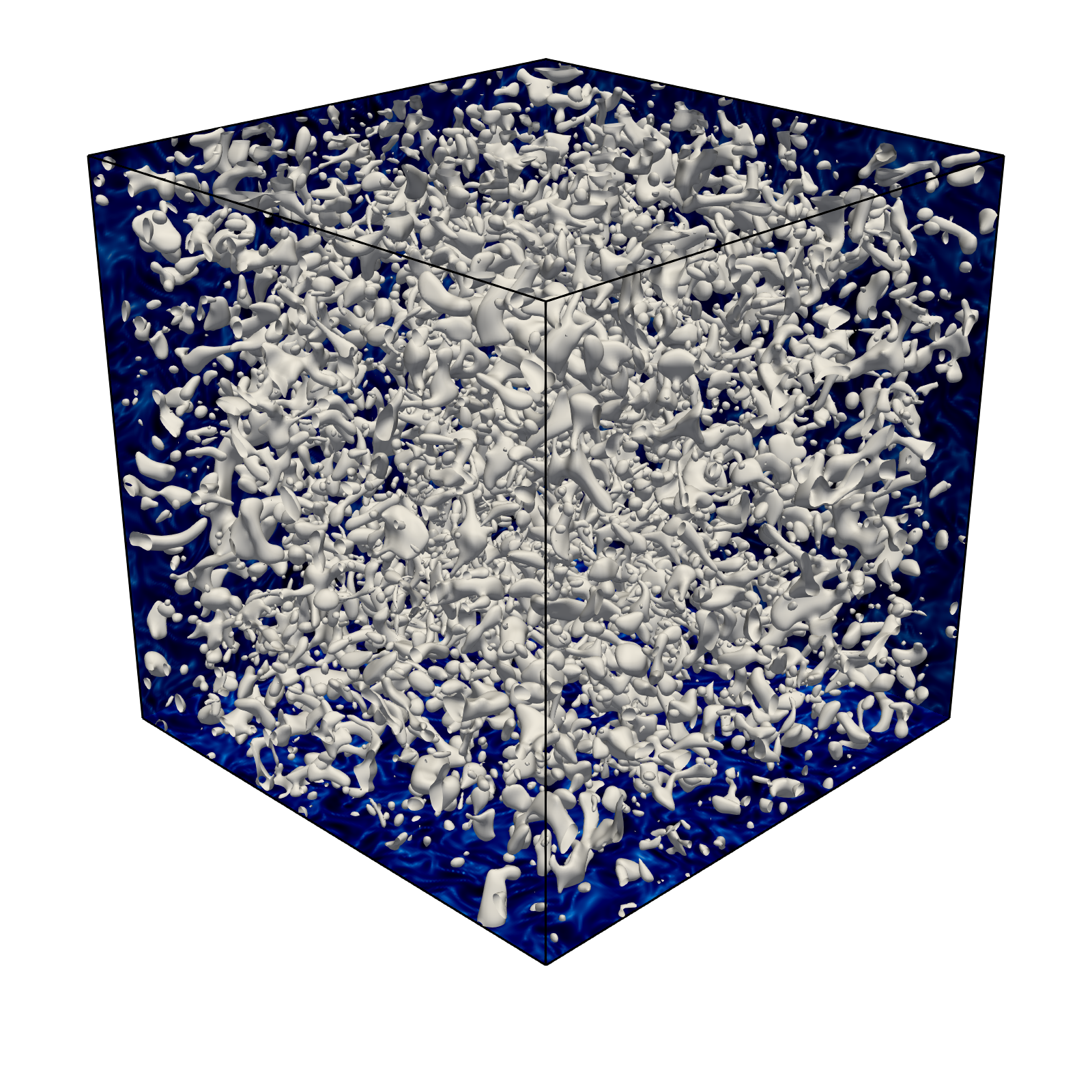}}; &
                \node[inner sep=0] (E2) {\includegraphics[width=0.24\textwidth]{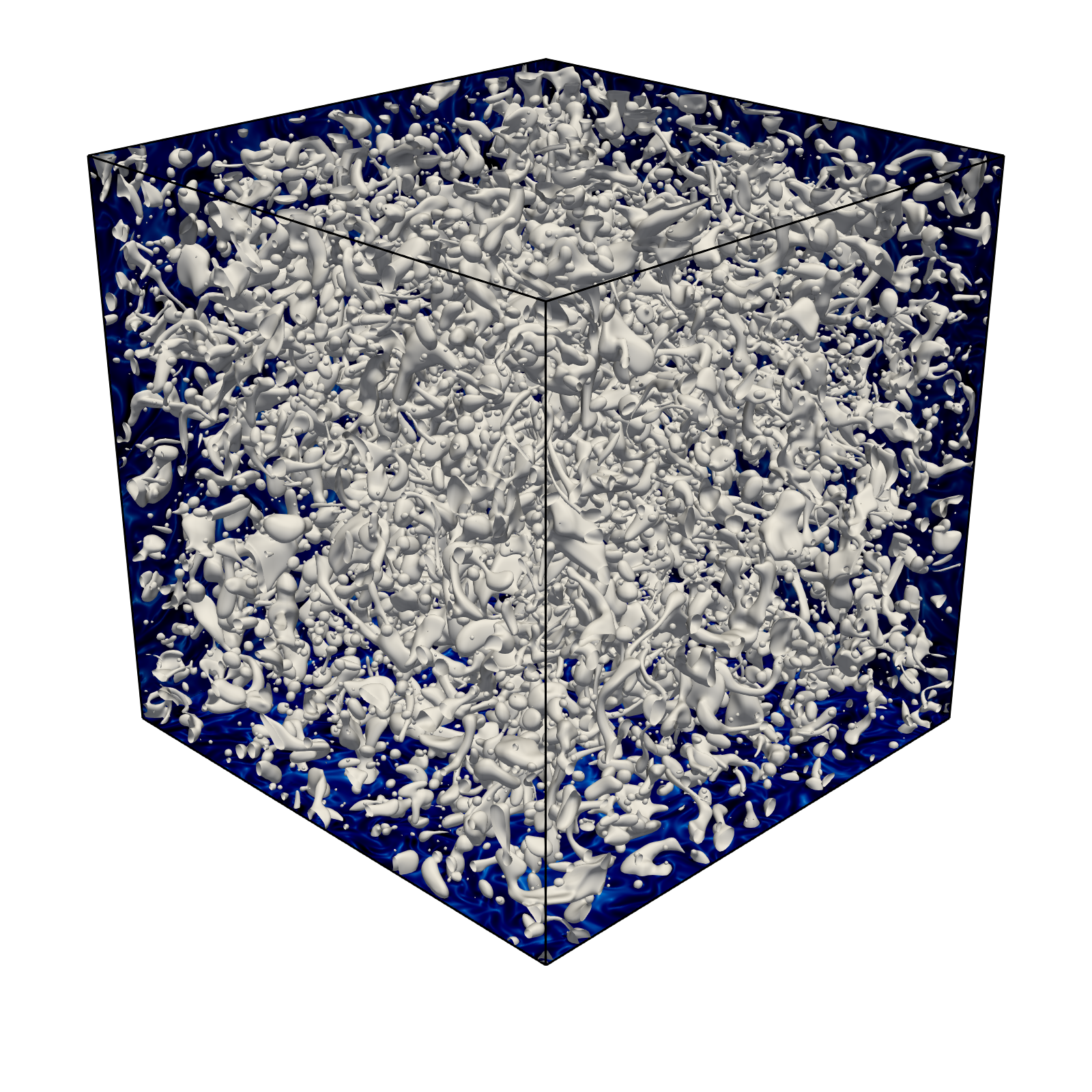}}; &
                \node[inner sep=0] (E3) {\includegraphics[width=0.24\textwidth]{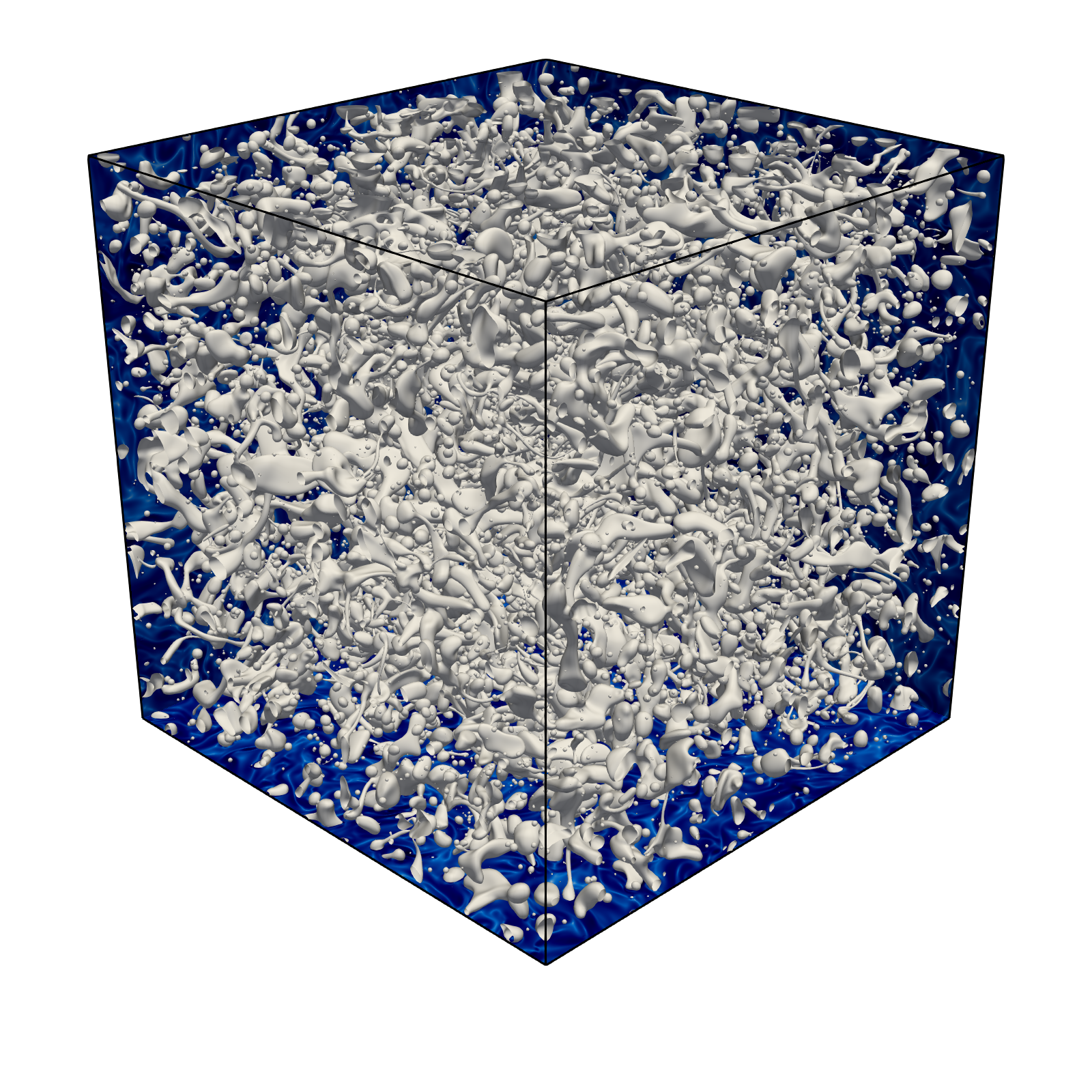}}; \\
            };
            % One rectangle around E2+E3
            \draw[blue, rounded corners, very thick] (E2.south west) rectangle (E3.north east);
        \end{tikzpicture}

        % Captions under each figure
        \parbox{0.24\textwidth}{\centering $k_{\max}\eta_{KH} = 15$}
        \parbox{0.24\textwidth}{\centering $k_{\max}\eta_{KH} = 30$}
        \parbox{0.24\textwidth}{\centering $k_{\max}\eta_{KH} = 60$}
        \parbox{0.24\textwidth}{\centering $k_{\max}\eta_{KH} = 120$}
    \end{minipage}

    \caption{Three-dimensional visualizations of instantaneous volume fraction contours, $\phi = 0.5$, over slices of enstrophy, $\Omega = \omega_i \omega_i$, at $t/\tau_e = 10$ for different grid resolutions. Cases A (top row) and B (bottom row). Solid blue line rectangles indicate the expected converged resolutions for both cases.}
    \label{fig:visualizations_3d_tau_e_10}
\end{figure}
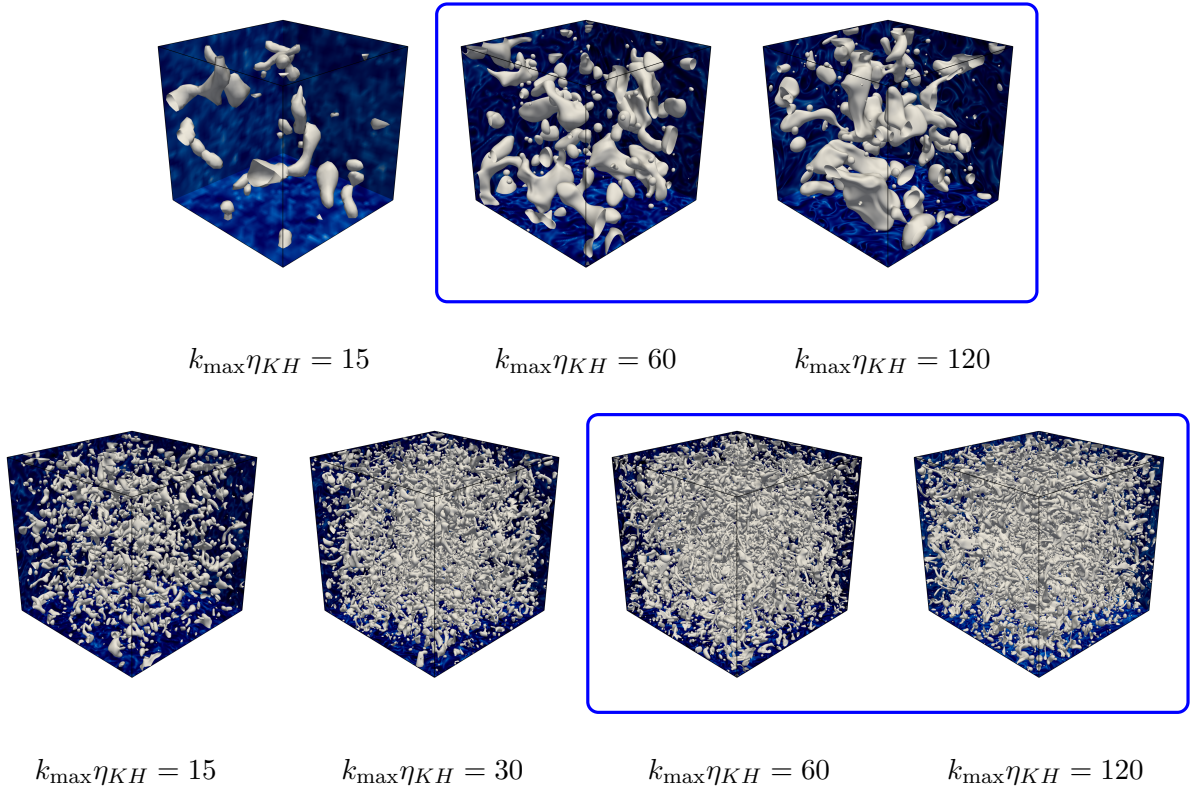

% b) Size distribution and SMD
\subsubsection{Size distribution}
\label{subsubsec:size_dist}

In the stationary state, the size distribution, $n(d)$, describes how the interfacial area is divided along different scales by determining the number of bubbles/droplets of size $d$. Although limited to spherical and ellipsoidal shapes, it can accurately compute the contribution of different self-contained scales in mass, momentum, and energy transfer phenomena. In the inertial regime, super-Kolmogorov-Hinze scales, $d\gtrsim \eta_{KH}$, cannot withstand pressure fluctuations due to turbulence, yielding a fragmentation cascade at least up to the Kolmogorov-Hinze scale~\cite{crialesi:2023}. For these sizes, dimensional analysis assuming locality yields in the $n(d) \sim d^{-10/3}$ power-law scaling~\cite{garrett:2000}. Sub-Kolmogorov-Hinze scaling, $d\lesssim \eta_{KH}$, does not have a well-established theoretical foundation, nevertheless, observations indicate that $n(d) \sim d^{-3/2}$~\cite{deane:2002,deike:2022,roccon2023phase,jain:2025}.

In addition to the convergence of the distributions themselves, the aforementioned scalings are used to facilitate grid resolution comparisons in Figure~\ref{fig:size_dist}. Here, size distributions are computed using the volume-corrected flood-fill algorithm that is known to account for the diffuse nature of the interface while computing the volumes and correcting for ellipticity, resulting in accurate size distributions~\cite{nathan:2025}. 
For the cases A in Figure~\ref{fig:size_dist} (a), the under-resolved simulations clearly overpredict super- and underpredict sub-Kolmogorov-Hinze scales. This is due to the fact that the grid cannot capture scales smaller than the grid size itself without the use of a subgrid model.
%, negatively skewing the distribution. 
Furthermore, the sub-Kolmogorov-Hinze scaling is also poorly captured. On the other hand, for the grid resolutions of $k_{\max} \eta_{KH} = 60$ and $120$, the distributions are converged over a minimum of a decade, whilst predicting both super- and sub-Kolmogorov-Hinze scalings with a minor mismatch for the smallest bubbles/droplets. 

\ifstrictfloat
\begin{figure}[H]
\else
\begin{figure}%[H]
\fi
    \centering
    \begin{subfigure}{0.495\linewidth}
        \centering
        \includegraphics[width=\linewidth]{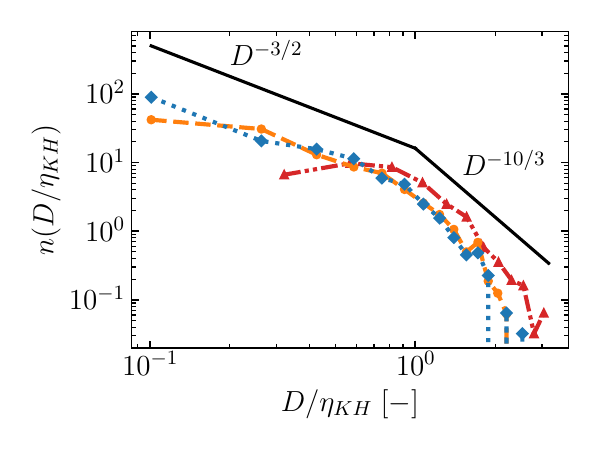}
        \caption{}
    \end{subfigure}
    \begin{subfigure}{0.495\linewidth}
        \centering
        \includegraphics[width=\linewidth]{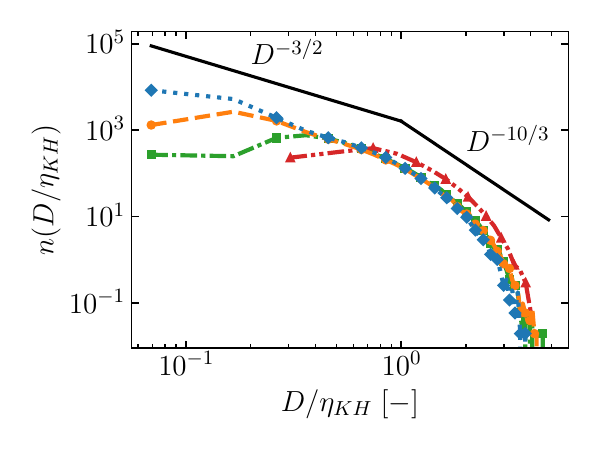}
        \caption{}
    \end{subfigure}
    \caption{Size distribution at steady state of (a) cases A, $Re_\lambda = 55$, $We_\mathcal{L} = 6.5$, (b) cases B, $Re_\lambda = 87$, $We_\mathcal{L} = 60$. Lines represent the measured size distribution for different grid resolutions. Solid black lines represent sub- and super-Kolmogorov-Hinze power law scalings. For the under-resolved cases, the number of bins, $n_h$, is chosen according to the Sturges' rule, where $n_h = \lceil \log_2 (n) \rceil + 1$ and $n$ is the number of samples, \textit{e.g.}, number of bubbles/droplets. For the cases with $k_{\max} \eta \geq 1.5$, the number of bins is computed as the average between the values obtained from the Sturges' and Rice's rule, where $n_h = \lceil 2 n^{1/3} \rceil$, to avoid over-smoothing and oscillations in the size distributions, respectively.}
    \label{fig:size_dist}
\end{figure}

For increased $Re_\lambda$ and $We_\mathcal{L}$, the range of scales is expected to increase, yielding in much finer turbulence scales and much smaller bubbles/droplets. For instance, compare for the same grid resolution in the Figure~\ref{fig:visualizations_3d_tau_e_10} the differences in the cases A and B. In the Figure~\ref{fig:size_dist} (b), the range of interface scales space is wider if compared to the Figure~\ref{fig:size_dist} (a), which is due to the presence of much smaller sub-Kolmogorov-Hinze bubbles/droplets. Similarly to the cases A, the under-resolved simulation of the cases B, $k_{\max} \eta_{KH} = 15$, fails to predict the $d \lesssim \eta_{KH}$ regime and its scaling. Within the super- and first half decade of the sub-Kolmogorov-Hinze regimes, all three grid resolutions starting from $k_{\max} \eta_{KH} = 30$ are converged. The deviation from convergence is more prominent only for the very small bubbles/droplets, where the number density increases with grid resolution. 
It is likely that intermittency, which has a greater turbulence resolution requirement, is playing a role in the breakup, therefore also yielding in the greater number density of very small sub-Kolmogorov-Hinze bubbles/droplets for $k_{\max} \eta_{KH} = 120$ (see, Appendix~\ref{apx:intermittency_area}). 
Comparing the scalings in the $d \lesssim \eta_{KH}$ regime, $k_{\max} \eta_{KH} = 60$ is seen to be sufficient except for the drops at the smallest scale.
%The finest resolution seems to enable the presence of intermittent events which yield in an increase of number density of very small bubbles/droplets. Furthermore, the coalescence time at these scales appears to be rather large\todo[]{I'm not sure if you can claim this just from area. I think this is too speculative, I'm not sure if we should comment on coalescence.} considering the total area increase from $10$ to $15\tau_e$ in the Figure~\ref{fig:area_evolution_cases_E}\todo[inline]{any reference on coalescence of very small sub-hinze? small colliding with small should be rare, but what is the coalesnce efficiency of collision between small and other sizes?}. For sufficiently large time averaging, the difference in number density at the smallest scales is likely to decrease and the convergence is expected to be better between $k_{\max} \eta_{KH} = 60$ and $120$. \todo[inline]{i would guess the reviewers would like us to run the simulation for more time and show the results, they won't take the comment for granted}
%\todo[]{Let's wait for the running simulations to complete and we can update the discussion according here.}

Although there is some discrepancy in the size distribution at the smallest scale, we can evaluate whether the contribution of those scales to the total interfacial area is significant or not, thereby indicating how substantial the impact on interfacial exchange phenomena would be. By definition, the interfacial area for a specific scale, $a(D)$, may be computed based on the number of bubbles/droplets obtained from the size distribution $n(D)$ as follows
\begin{equation}
    \label{eq:area_contribution}
    a(D) = n(D) \pi D^2.
\end{equation}
The total interfacial area is the summation of the interfacial area contribution of all scales,
\begin{equation}
    \label{eq:total_area_from_contribution}
    A = \sum_{D_{\min}}^{D_{\max}} a(D) = \sum_{D_{\min}}^{D_{\max}} n(D) \pi D^2.
\end{equation}
The Figure~\ref{fig:area_contribution_cases_E} depicts the percentage of the interfacial area contributed by each scale to the total interfacial area for the cases B, $Re_\lambda = 87$ and $We_\mathcal{L} = 60$. Based on the size distribution power law scalings for sub- and super-Kolmogorov-Hinze scales, one can also derive the interfacial area scalings for both regimes by considering the additional $D^2$ contribution from the surface area. Both sub- and super-Kolmogorov-Hinze area power laws are plotted for reference, which are $a(D<\eta_{KH}) \sim D^{1/2}$ and $a(D>\eta_{KH}) \sim D^{-4/3}$, respectively. Among the turbulence-resolved cases, it is clearer that $k_{\max} \eta_{KH} = 30$ is not sufficient to capture the sub-Kolmogorov-Hinze area scaling. When the grid resolution is increased to $k_{\max} \eta_{KH} = 60$, the scaling is improved. Furthermore, considering the finest grid resolution, the area contribution of the smallest scale corresponds to only $1\%$ of the total interfacial area. Therefore, even if there is some discrepancy at those scales, the differences in total interfacial area are not significant and interface exchange phenomena, such as phase and heat change, could still be properly captured by the proposed resolution of $k_{\max} \eta_{KH} = 60$. Super-Kolmogorov-Hinze area contribution scaling is properly predicted for all turbulence-resolved cases, given that all grids can properly represent large scales discretely.

%\todo[inline]{should i comment on how size distribution is limited and cannot capture all area? currently the size distribution captures approximately 80\% of the total interfacial area}\todo[inline]{probably not needed, I dont know if the remaining 20\% is real or not. They could be from spurious drops.}

\ifstrictfloat
\begin{figure}[H]
\else
\begin{figure}%[H]
\fi
    \centering
    \includegraphics[width=0.5\linewidth]{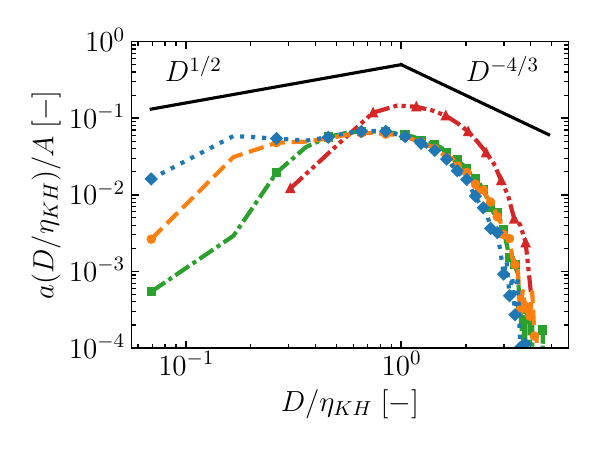}
    \caption{Area percentage contribution to total interfacial by each scale of cases B, $Re_\lambda = 87$, $We_\mathcal{L} = 60$, for different grid resolutions. Based on integration of size distribution power law scalings, area scalings for sub- and super-Kolmogorov-Hinze scales are added as solid black lines.}
    \label{fig:area_contribution_cases_E}
\end{figure}

\subsubsection{Sauter mean diameter}
\label{subsubsec:smd}

The Sauter mean diameter is a single reference scalar of the size distribution. By definition, it is also a function of the second and third moments of the distribution, thus, super-Kolmogorov-Hinze bubbles/droplets should have a higher contribution than the sub-Kolmogorov-Hinze ones. In Figure~\ref{fig:comp_avg_smd}, for the set of grid resolutions, the convergence of SMD is depicted for both pairs of $Re_\lambda$ and $We_\mathcal{L}$. Considering one standard deviation, for both cases A and B, all SMDs with grid resolution above $k_{\max} \eta_{KH} = 30$ are converged, mostly due to the contribution of larger bubbles/droplets in higher-order moments of the size distribution.

\ifstrictfloat
\begin{figure}[H]
\else
\begin{figure}%[H]
\fi
    \centering
    \includegraphics[width=0.65\linewidth]{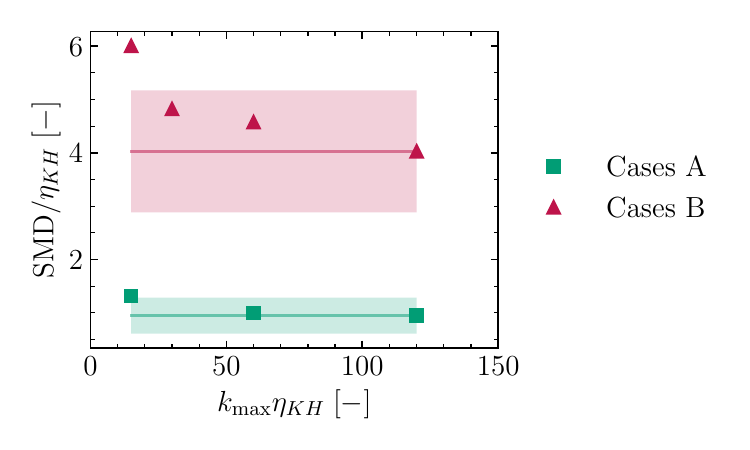}
    \caption{Sauter mean diameter at steady state comparison for two pairs of dimensionless number as a function of grid resolution, where $Re_\lambda = 55$, $We_\mathcal{L} = 6.5$ for the cases A, and, $Re_\lambda = 87$, $We_\mathcal{L} = 60$ for the cases E. Solid line based on the finest resolution value and one standard deviation bands added for reference.}
    \label{fig:comp_avg_smd}
\end{figure}

% c) Curvature distribution
\subsubsection{Interface curvature}
\label{subsubsec:curv}

The interface curvature characterizes the local deformation of bubbles and droplets and is also a well-defined quantity in transient states of atomization, indicating likely regions of breakup. Considering an outward normal for the dispersed phase, positive values represent convex surfaces, zero values flat surfaces, and negative values concave surfaces. Furthermore, the curvature distribution indicates the probability of deviations from the spherical equilibrium shape, highly positive values of curvature represent small bumps and ripples on the surface and highly negative values small dimples. These features can be characterized as corrugations on the interface and can also contribute to the overall interfacial area~\cite{mangani:2022}. In the proposed setup of stationary HIT, the dispersed phase is expected to be composed of bubbles/droplets of spherical and ellipsoidal shapes with corrugations for those of larger scales. 

The Figure~\ref{fig:pdf_curv} shows the probability density functions (PDF) of the normalized interface curvature for cases A and B for varying grid resolutions. Qualitatively, for all cases and grid resolutions, the distributions are positively skewed as expected given the on-average convex nature of the dispersed phase. Nevertheless, there are significant differences with respect to the likelihoods of high-magnitude curvature values, corresponding to interface corrugations (both positive and negative curvatures) and very small bubbles and droplets (mostly positive curvature). In order to quantify the discrepancies in likelihood, horizontal lines are drawn to determine the curvature bounds for specified percentages of the definite integral of the PDF within these bounds, as done in~\cite{yang:2021b}. Lines for percentages of 90\%, 95\%, 99\%, and 99.9\% present different levels of fidelity of the described quantity based on the finest grid resolution, $k_{\max} \eta_{KH} = 120$. For both cases, there is a distinction on how accurately negative and positive values of curvature are captured. For instance, negative values, representative of surface dimples, are well captured almost up to 99.9\% for the $k_{\max} \eta_{KH}=60$ resolution. However, for positive values, representative of bumps and bubbles and droplets themselves, larger-magnitude instances are not captured to the same extent. In Figure~\ref{fig:pdf_curv} (a), positive values of curvature for $k_{\max} \eta_{KH}=60$ are well-captured up to 95\%, while in Figure~\ref{fig:pdf_curv} (b), they are approximately well-captured up to 99\%. 
%One may suppose that both convex and concave corrugation scales are of the same order, thus, if the concave ones are properly captured, then so are the convex ones. 
The discrepancy at very large values of positive curvature is probably due to the small amount of tiny bubbles and droplets at the finest grid, which typically do not contribute significantly to the interfacial area. 
%For instance, refer again to the Figure~\ref{fig:visualizations_slices_cases_E}, at $t/\tau_e=15$, there is a more significant amount of very small bubbles and droplets comparing $k_{\max} \eta_{KH} = 120$ to $60$.

\ifstrictfloat
\begin{figure}[H]
\else
\begin{figure}%[H]
\fi
    \centering
    \begin{subfigure}{0.495\linewidth}
        \centering
        \includegraphics[width=\linewidth]{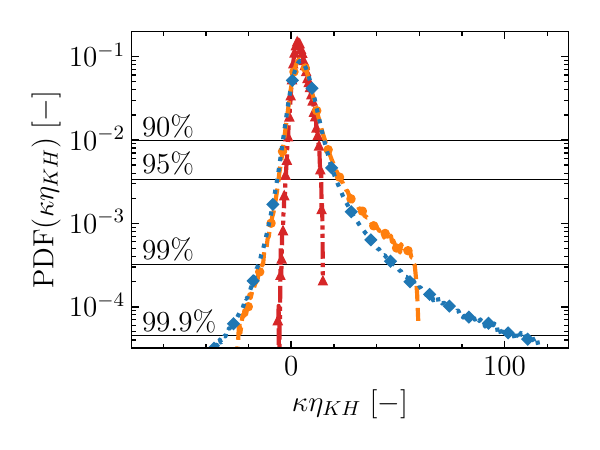}
        \caption{}
    \end{subfigure}
    \begin{subfigure}{0.495\linewidth}
        \centering
        \includegraphics[width=\linewidth]{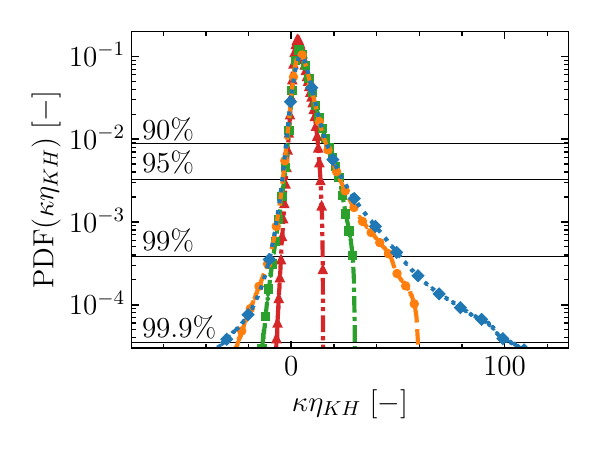}
        \caption{}
    \end{subfigure}
    \caption{Curvature distribution at steady state of (a) cases A, $Re_\lambda = 55$, $We_\mathcal{L} = 6.5$, (b) cases E, $Re_\lambda = 87$, $We_\mathcal{L} = 60$. Horizontal lines represent the percentage of the definite integral of the PDF captured by the bounds at each cutoff.}
    \label{fig:pdf_curv}
\end{figure}

\subsection{Discussion}
\label{subsec:disc}

% in Yang and Griffin's discussion, they mention:
% - how their estimates compare with what W. C. Reynolds had proposed;
% - what would the viscous time-step requirement imply in the cost;

% I could talk about:
% - make estimates of how would the cost increase with We number;
% - reiterate the how resolution should be interpreted with respect to other schemes;

Similar to the estimates in terms of grid-points and time-step requirements proposed by W. C. Reynolds~\cite{reynolds:1990}, two-phase flow scalings derived in Section~\ref{sec:scalings} can be used to provide an outlook of the future of high-fidelity multiphase simulations. In single-phase flows, an increase by a factor of $10$ in $Re_{\mathcal{L}}$ requires approximately 180 times increase in the number of grid points and 5 times in the number of time steps, therefore, the corresponding cost, roughly computed as $N*N_t$~\cite{reynolds:1990}, is increased by 900 times, almost 2 orders of magnitude difference in the increase of $Re_{\mathcal{L}}$ and cost. For two-phase flows, in the inertia-dominated turbulent regime, an increase by a factor of $10$ in $We_{\mathcal{L}}$ requires approximately $60$ times more grid points and 4 times more time steps, thus, 240 times greater cost. For the viscous-dominated regime, given the $\mathcal{L}/\eta_{KV} \sim Ca_{\mathcal{L}}$, for an increase of 10 times in $Ca_{\mathcal{L}}$, the number of grid points is increased by 1000 times and the number of time steps by 10 times, resulting in a significant increase in cost of 10000 times.

So far the verification studies were limited to the inertia-dominated regime due to the challenges with the very high resolution requirements and the computational power constraints to achieve the viscous-dominated regime whilst having a sufficiently large range of turbulent scales. In order to achieve such regime, surface tension has to be significantly reduced such that $\eta_{KV} \ll \eta_{KH}$, \textit{i.e.}, $Ris \gg 1$, which requires $We_{\mathcal{L}}>O(10^8)$ for $Re_\lambda=87$. Therefore, if one assumes $k_{\max} \eta_{KH} \sim k_{\max}\eta_{KV}$, then, for the current numerical methods, one would require $N_x \sim 2^{25}=O(10^{7})$, thus, $N \sim 2^{75}=O(10^{22})$ or over Sextillion grid points.

In the inertia-dominated regime, considering the current results for the cases B with $Re_\lambda=87$ and $We_{\mathcal{L}}=60$, a grid-resolution requirement of $k_{\max} \eta_{KH}=60$ was observed for the adopted low-dissipative second-order central scheme. First of all, as a remark, this value should not be considered as a strict bound for all other numerical methods, it should serve as a reference point. For instance, in the case of high-order schemes, the grid resolution could be slightly reduced. Second, based on the observed grid resolution requirement and the set of dimensionless numbers, estimates on the computation demand to achieve realistic regimes for real life applications can be deduced. Here, a grid of $512^3$ was required for $We_{\mathcal{L}}=60$ and $k_{\max}\eta_{KH}=60$. As two examples, consider breaking waves and fuel injection, the corresponding Weber numbers would be on the order of $10^4$~\cite{iafrati:2009,wang:2016,chan:2021,digiorgio:2022} and $10^3 - 10^4$~\cite{lubarsky:2009,liu:2022}, respectively. Therefore, assuming only the increase in $We_{\mathcal{L}}$, from Eq.~\eqref{eq:total_grid_surf_wel}, 
%realistic raindrop formation could be studied with approximately 158 times more grid points, roughly a grid of $2760^3$. Moreover, 
realistic breaking waves and fuel injection would require from 158 to 10000 times more grid points, which translates to $2760^3$ and $11025^3$ grids, both of which would be currently achievable considering the boundaries realized for the single-phase turbulence simulations of $32768^3$~\cite{yeung:2023}. Overall, a promising outlook. 

\section{Conclusions}
\label{sec:conc}

% - main outcome is phase plot with all regimes and the corresponding approximate resolution requirements;

% outline:
% - recall scalings;
% - recall main grid and time step requirements;
% - talk about verification;

In this work, important two-phase flow scalings are derived in both inertia- and viscous-dominated regimes. Such results allow for the extension of grid-point and time-step cost estimates previously only limited to single-phase flows. For the inertia-dominated regime, the interface restricts the total number of grid points, $N$, to scale with $We_{\mathcal{L}}^{9/5}$ and the total number of time steps, $N_t$, when limited by the convective CFL, to scale with $We_{\mathcal{L}}^{3/5}$, thus, the computational cost $N * N_t$ approximately scales with $We_{\mathcal{L}}^{12/5}$. For the viscous-dominated regime, the capillary interface scale imposes $N \sim Ca_{\mathcal{L}}^3$ and $N_t \sim {Ca}_{\mathcal{L}}$, and the corresponding cost $N*N_t \sim Ca_{\mathcal{L}}^4$. 

In order to provide a clearer delimitation of the inertia- and viscous-dominated regimes, a new non-dimensional number is proposed, denoted as the \textit{ratio of interface scales} ($Ris$). It is defined as the ratio of the Kolmogorov-Hinze to the Kolmogorov-viscous scales, $\eta_{KH}/\eta_{KV}$, which can be also expressed as $Ris = We_\mathcal{L}^{2/5}/Re_\mathcal{L}$. The smallest of the scales is the one to govern breakup at smallest bubbles/droplets. By definition, the inertia-dominated regime is characterized by $Ris \ll 1$ and the viscous-dominated regime by $Ris \gg 1$.

In addition to the scalings, numerical simulations are performed to determine the grid resolution requirements in the inertia-dominated regime. Since, it is currently not achievable to perform resolved simulations in the viscous-dominated regime based on rough estimates of $N_{KV}\sim O(10^{75})$ or Sextillion grid points.
%, which is approximately 1 billion times bigger than one of the currently largest simulations ($32768^3$ grid points)~\cite{yeung:2023}. 
For two pairs of $Re_\lambda$ and $We_{\mathcal{L}}$, the grid resolution $k_{\max} \eta_{KH}$ is varied and quantities of interest, such as total interfacial area, size distribution, SMD, and curvature PDF, are evaluated for convergence. Considering a low-dissipative second-order scheme and ACDI as the interface-capturing method, the results demonstrate $k_{\max} \eta_{KH}=60$ to be sufficient to capture the aforementioned quantities. As $Re_\lambda$ increases and extreme events increase in magnitude and frequency, very small bubbles/droplets, an order of magnitude smaller than $\eta_{KH}$, are more likely to be formed. For the finest resolution evaluated herein of $k_{\max} \eta_{KH}=120$, the size distribution presents a greater number density of those scales if compared to the other resolutions. Scale by scale total interfacial area percentages are computed and the corresponding contribution from tiny bubbles/droplets is found to be approximately $1\%$, hence, not contributing enough to bulk exchanges of mass/momentum/energy between the phases. 
%Although marginally underpredicting the number density of the smallest interface scales, 
The proposed resolution is also sufficient to capture intermittent states of increased energy dissipation and therefore increased number of small bubbles/droplets.

As a final remark, we expect the current results to provide information and guidelines on the future of two-phase numerical simulations. 
Resolution requirements should be intelligently translated to other methodologies, \textit{e.g.}, reduce $k_{\max} \eta_{KH}$ for high-order methods. 
Furthermore, for the inertia-dominated regime, the future seems more promising in achieving more realistic levels of $We$. On the other hand, for the viscous-dominated regime, computational power is currently not sufficient.

\begin{acknowledgments}
Authors acknowledge support from the George W. Woodruff School of Mechanical Engineering at Georgia Institute of Technology. 
P.~J.~N. acknowledges financial support from the President's Undergraduate Research Award (PURA) at Georgia Institute of Technology and S.~S.~J acknowledges support by the donors of ACS Petroleum Research Fund under Doctoral New Investigator Grant 69196-DNI9 (S.~S.~J. served as Principal Investigator on ACS PRF 69196-DNI9). Authors also acknowledge the generous computing resources from the DOE's 2024 and 2025 ALCC awards (TUR147 \& BubbleLaden, PI: Jain). This research used supporting resources at the Argonne and the Oak Ridge Leadership Computing Facilities. The Argonne Leadership Computing Facility at Argonne National Laboratory is supported by the Office of Science of the U.S. DOE under Contract No. DE-AC02-06CH11357. The Oak Ridge Leadership Computing Facility at the Oak Ridge National Laboratory is supported by the Office of Science of the U.S. DOE under Contract No. DE-AC05-00OR22725.
\end{acknowledgments}

\appendix

\section{Requirements for minimum step size and total number of iterations}
\label{apx:time_step_reqs_complete}

The time-step size, $dt$, can be normalized by $\nu/u'^2$ in the Eq.~\eqref{eq:dts_1}, which will be used in determining the total number of time steps for converged statistics.
Consider first the viscous-restricted time step based on the small turbulence length scale, $\Delta_{\rm turb}$, constraint,
\begin{eqnarray}
    \label{eq:dt_visc_kmax_eta}
    \frac{u'^2 dt}{\nu} &\leq& \frac{\Delta_{\rm turb}^2}{\nu} \frac{u'^2}{\nu} = \left ( \frac{\pi}{k_{\max} \eta} \right )^2 \frac{\eta^2}{\mathcal{L}^2} \frac{u'^2 \mathcal{L}^2}{\nu^2}, \nonumber\\
    &\leq& \frac{2}{3} \left ( \frac{\pi}{k_{\max} \eta} \right )^2 Re_\mathcal{L}^{1/2} =  \frac{\sqrt{15}}{15} \left ( \frac{\pi}{k_{\max} \eta} \right )^2 Re_\lambda.
\end{eqnarray}
Moreover, with respect to the inertia-dominated interface length scale, $\Delta_{\rm surf}$, constraint,
\begin{eqnarray}
    \label{eq:dt_visc_kmax_eta_kh}
    \frac{u'^2 dt}{\nu} &\leq& \frac{\Delta_{\rm surf}^2}{\nu} \frac{u'^2}{\nu} = \left ( \frac{\pi}{k_{\max} \eta_{KH}} \right )^2 \frac{\eta_{KH}^2}{\mathcal{L}^2} \frac{u'^2 \mathcal{L}^2}{\nu^2}, \nonumber\\
    &\leq& { \left (\frac{2}{3} \right )^{11/5}} C_2^{-4/5} \left ( \frac{\pi}{k_{\max} \eta_{KH}} \right )^2 \left (\frac{We_t^c}{We_\mathcal{L}} \right )^{6/5} Re_\mathcal{L}^2, \nonumber \\
    &\leq& { \frac{1}{15^{4/5}}} C_2^{-4/5} \left ( \frac{\pi}{k_{\max} \eta_{KH}} \right )^2 \left (\frac{We_t^c}{We_\lambda} \right )^{6/5} Re_\lambda^{14/5}.
\end{eqnarray}
Lastly, with respect to the capillary interface length scale, $\Delta_{\rm cap}$, constraint, 
\begin{eqnarray}
    \label{eq:dt_visc_kmax_eta_kv}
    \frac{u'^2 dt}{\nu} &\leq& \frac{\Delta_{\rm cap}^2}{\nu} \frac{u'^2}{\nu} = \left ( \frac{\pi}{k_{\max} \eta_{KV}} \right )^2 \frac{\eta_{KV}^2}{\mathcal{L}^2} \frac{u'^2 \mathcal{L}^2}{\nu^2}, \nonumber\\
    &\leq& \frac{2}{3} \left ( \frac{\pi}{k_{\max} \eta_{KV}} \right )^2 \left (\frac{Ca_t^c}{Ca_\mathcal{L}} Re_\mathcal{L} \right )^2 = \left ( \frac{\pi}{k_{\max} \eta_{KV}} \right )^2 \left (\frac{Ca_t^c}{Ca_\lambda} Re_\lambda \right )^2.
\end{eqnarray}
Consider now the generic convective/interface-restricted time step as a function of the small turbulence resolution, $\Delta_{\rm turb}$, constraint.
\begin{eqnarray}
    \label{eq:dt_surf_kmax_eta} 
    \frac{u'^2 dt}{\nu} &\leq& \frac{1}{\alpha} \frac{\Delta_{\rm turb}}{|u_{\max} |} \frac{u'^2}{\nu} = \frac{1}{\alpha} \frac{\pi}{k_{\max} \eta} \frac{u'}{|u_{\max}|} \frac{\eta}{\mathcal{L}} \frac{u' \mathcal{L}}{\nu}, \nonumber\\
    &\leq& \sqrt{\frac{2}{3}} \frac{1}{\alpha} \frac{\pi}{k_{\max} \eta} \frac{u'}{|u_{\max}|} Re_\mathcal{L}^{1/4} = \left ( \frac{1}{15}\right )^{1/4} \frac{1}{\alpha} \frac{\pi}{k_{\max} \eta} \frac{u'}{|u_{\max}|} Re_\lambda^{1/2}.
\end{eqnarray}
Furthermore, with respect to the inertia-dominated interface length scale constraint, $\Delta_{\rm surf}$,
\begin{eqnarray}
    \label{eq:dt_surf_kmax_eta_kh} 
    \frac{u'^2 dt}{\nu} &\leq& \frac{1}{\alpha} \frac{\Delta_{\rm surf}}{|u_{\max} |} \frac{u'^2}{\nu} = \frac{1}{\alpha} \frac{\pi}{k_{\max} \eta_{KH}} \frac{u'}{|u_{\max}|}\frac{\eta_{KH}}{\mathcal{L}} \frac{u' \mathcal{L}}{\nu}, \nonumber\\
    &\leq& { \left ( \frac{2}{3} \right )^{11/10}} C_2^{-2/5} \frac{1}{\alpha} \frac{\pi}{k_{\max} \eta_{KH}} \frac{u'}{|u_{\max}|} \left ( \frac{We_t^c}{We_\mathcal{L}} \right )^{3/5} Re_\mathcal{L}, \nonumber\\
    &\leq& { \frac{1}{15^{2/5}}} C_2^{-2/5} \frac{1}{\alpha} \frac{\pi}{k_{\max} \eta_{KH}} \frac{u'}{|u_{\max}|} \left ( \frac{We_t^c}{We_\lambda} \right )^{3/5} Re_\lambda^{7/5}.
\end{eqnarray}
Lastly, with respect to the capillary interface length scale, $\Delta_{\rm cap}$, constraint,
\begin{eqnarray}
    \label{eq:dt_surf_kmax_eta_kv} 
    \frac{u'^2 dt}{\nu} &\leq& \frac{1}{\alpha} \frac{\Delta_{\rm cap}}{|u_{\max} |} \frac{u'^2}{\nu} = \frac{1}{\alpha} \frac{\pi}{k_{\max} \eta_{KV}} \frac{u'}{|u_{\max}|} \frac{\eta_{KV}}{\mathcal{L}} \frac{u' \mathcal{L}}{\nu}, \nonumber\\
    &\leq& \sqrt{\frac{2}{3}} \frac{1}{\alpha} \frac{\pi}{k_{\max} \eta_{KV}} \frac{u'}{|u_{\max}|} \frac{Ca_t^c}{Ca_\mathcal{L}} Re_\mathcal{L} = \frac{1}{\alpha} \frac{\pi}{k_{\max} \eta_{KV}} \frac{u'}{|u_{\max}|} \frac{Ca_t^c}{Ca_\lambda} Re_\lambda.
\end{eqnarray}

In summary, the normalized time-step size scalings as a function of the several reference length scale and restricting numerical integration stability bounds are presented in the Table~\ref{tab:summary_multi_turb_dt}. Considering only the power-law scalings, one may relate the convective/interface and viscous normalized time-step sizes for each reference length scale by $dt_{\rm visc} \sim dt_{\rm conv,surf}^2$, since the inverse of the CFLs scale as $1/{\rm CFL_{conv,surf}} \sim \Delta$ and $1/{\rm CFL_{visc}} \sim \Delta^2$.
\ifstrictfloat
\begin{table}[H]%% placement specifier
\else
\begin{table}%[H]%% placement specifier
\fi
\centering%% For centre alignment of tabular.
\caption{Power laws for normalized time-step sizes, $u'^2 dt/\nu$, in turbulence, inertia-dominated, and viscous-dominated regimes bounded by convective/interface and viscous stability criteria as functions of $Re$, $We$, $Ca$, and $\Delta_{\rm turb}$ ($k_{\max} \eta$), $\Delta_{\rm surf}$ ($k_{\max} \eta_{KH}$), and $\Delta_{\rm cap}$ ($k_{\max} \eta_{KV}$).}\label{tab:summary_multi_turb_dt}
\begin{tabular}{c | c | c}%% Table column specifiers
  \hline
                           & $dt_{\rm conv/surf}$                  & $dt_{\rm visc}$                         \\ \hline
   $\Delta_{\rm turb}$     & $Re_\mathcal{L}^{1/4}$                & $Re_{\mathcal{L}}^{1/2}$                \\
   $\Delta_{\rm surf}$     & $We_\mathcal{L}^{-3/5}Re_\mathcal{L}$ & $We_\mathcal{L}^{-6/5}Re_\mathcal{L}^2$ \\
   $\Delta_{\rm cap}$      & $Ca_\mathcal{L}^{-1}Re_\mathcal{L}$   & $Ca_\mathcal{L}^{-1}Re_\mathcal{L}$     \\ \hline
\end{tabular}
\end{table}

The total number of time steps is computed based on minimum time-step sizes previously defined for all combinations of numerical stability bounds and grid size resolutions due to different physics. Previously derived results for $u'^2 dt / \nu$ are substituted in the Eq.~\eqref{eq:nt}. First, consider the viscous CFL-limited total number of time steps. From Eqs.~\eqref{eq:dt_visc_kmax_eta},~\eqref{eq:dt_visc_kmax_eta_kh},~\eqref{eq:dt_visc_kmax_eta_kv}, and~\eqref{eq:nt}, for the carrier turbulence-, two-phase inertia-dominated-, and viscous-dominated-limited grid size resolutions, it follows $N_t$ respectively, 
\begin{eqnarray}
    \label{eq:nt_visc_eta}
    N_{t, \rm visc, turb} &\geq& \left . \sqrt{\frac{2}{3}} C_t Re_\mathcal{L} \middle / \frac{2}{3} \left (\frac{\pi}{k_{\max} \eta} \right )^2 Re_\mathcal{L}^{1/2} \right . \nonumber \\
    &\geq& \sqrt{\frac{3}{2}} C_t \left (\frac{k_{\max} \eta}{\pi} \right)^2 Re_\mathcal{L}^{1/2} = \frac{3\sqrt{10}}{20} C_t \left (\frac{k_{\max} \eta}{\pi} \right)^2 Re_\lambda,
\end{eqnarray}
\begin{eqnarray}
    \label{eq:nt_visc_eta_kh}
    N_{t, \rm visc, surf} &\geq& \left . \sqrt{\frac{2}{3}} C_t Re_\mathcal{L} \middle / \left (\frac{2}{3} \right )^{11/5} C_2^{-4/5}\left (\frac{\pi}{k_{\max} \eta_{KH}} \right )^2 \fpp{\frac{We_t^c}{We_\mathcal{L}}}{6/5} Re_\mathcal{L}^{2} \right . \nonumber \\
    &\geq& { \fpp{\frac{3}{2}}{17/10}} C_2^{4/5} C_t \left (\frac{k_{\max} \eta_{KH}}{\pi} \right)^2 \fpp{\frac{We_\mathcal{L}}{We_t^c}}{6/5} Re_\mathcal{L}^{-1} \nonumber \\
    &\geq& { \frac{\sqrt{6}}{20}}15^{4/5} C_2^{4/5} C_t \left (\frac{k_{\max} \eta_{KH}}{\pi} \right )^2 \fpp{\frac{We_\lambda}{We_t^c}}{6/5} Re_\lambda^{-4/5},
\end{eqnarray}
\begin{eqnarray}
    \label{eq:nt_visc_eta_kv}
    N_{t, \rm visc, cap} &\geq& \left . \sqrt{\frac{2}{3}} C_t Re_\mathcal{L} \middle / \frac{2}{3} \left (\frac{\pi}{k_{\max} \eta_{KV}} \right )^2 \left (\frac{Ca_t^c}{Ca_\mathcal{L}} \right )^2 Re_\mathcal{L}^{2} \right . \nonumber \\
    &\geq& \sqrt{\frac{3}{2}} C_t \left (\frac{k_{\max} \eta_{KV}}{\pi} \right)^2 \left ( \frac{Ca_\mathcal{L}}{Ca_t^c} \right )^2 Re_\mathcal{L}^{-1} = \frac{\sqrt{6}}{20} C_t \left (\frac{k_{\max} \eta_{KV}}{\pi} \right)^2 \left ( \frac{Ca_\lambda}{Ca_t^c} \right )^2.
\end{eqnarray}
Second, consider the convective/interface CFL-restricted total number of time steps. From Eqs.~\eqref{eq:dt_surf_kmax_eta},~\eqref{eq:dt_surf_kmax_eta_kh},~\eqref{eq:dt_surf_kmax_eta_kv} and~\eqref{eq:nt}, again for the carrier turbulence-, two-phase inertia-dominated-, and viscous-dominated-limited grid size resolutions, it follows $N_t$ respectively,
\begin{eqnarray}
    \label{eq:nt_surf_eta}
    N_{t, \rm conv, turb} &\geq& \left . \sqrt{\frac{2}{3}} C_t Re_\mathcal{L} \middle / \sqrt{\frac{2}{3}} \frac{1}{\alpha} \frac{\pi}{k_{\max} \eta} \frac{u'}{|u_{\max}|} Re_\mathcal{L}^{1/4} \right . \nonumber \\
    &\geq& C_t \alpha \frac{k_{\max} \eta}{\pi} \frac{|u_{\max}|}{u'} Re_\mathcal{L}^{3/4} = \fpp{\frac{3}{20}}{3/4} C_t \alpha \frac{k_{\max} \eta}{\pi} \frac{|u_{\max}|}{u'} Re_\lambda^{3/2},
\end{eqnarray}
\begin{eqnarray}
    \label{eq:nt_surf_eta_kh}
    N_{t, \rm conv, surf} &\geq& \left . \sqrt{\frac{2}{3}} C_t Re_\mathcal{L} \middle / \left (\frac{2}{3} \right )^{11/10} C_2^{-2/5} \frac{1}{\alpha} \frac{\pi}{k_{\max} \eta_{KH}} \frac{u'}{|u_{\max}|} \fpp{\frac{We_t^c}{We_\mathcal{L}}}{3/5} Re_\mathcal{L} \right . \nonumber \\
    &\geq& { \left (\frac{3}{2} \right )^{3/5}} C_t C_2^{2/5} \alpha \frac{k_{\max} \eta_{KH}}{\pi} \frac{|u_{\max}|}{u'} \fpp{\frac{We_\mathcal{L}}{We_t^c}}{3/5} \nonumber \\
    &\geq& { \left (\frac{3}{2} \right )^{3/5} \fpp{\frac{3}{200}}{3/10}} C_t C_2^{2/5} \alpha \frac{k_{\max} \eta_{KH}}{\pi} \frac{|u_{\max}|}{u'} \fpp{\frac{We_\lambda }{We_t^c}Re_\lambda}{3/5},
\end{eqnarray}
\begin{eqnarray}
    \label{eq:nt_surf_eta_kv}
    N_{t, \rm conv, cap} &\geq& \left . \sqrt{\frac{2}{3}} C_t Re_\mathcal{L} \middle / \sqrt{\frac{2}{3}} \frac{1}{\alpha} \frac{\pi}{k_{\max} \eta_{KV}} \frac{u'}{|u_{\max}|} \frac{Ca_t^c}{Ca_\mathcal{L}} Re_\mathcal{L} \right . \nonumber \\
    &\geq& C_t \alpha \frac{k_{\max} \eta_{KV}}{\pi} \frac{|u_{\max}|}{u'} \frac{Ca_\mathcal{L}}{Ca_t^c} = \frac{\sqrt{6}}{20} C_t \alpha \frac{k_{\max} \eta_{KV}}{\pi} \frac{|u_{\max}|}{u'} \frac{Ca_\lambda}{Ca_t^c} Re_\lambda.
\end{eqnarray}
The summary of the total number of time steps scalings are presented in the Table~\ref{tab:summary_multi_turb_nt}.

\section{Effect of intermittency on interfacial area}
\label{apx:intermittency_area}

Considering the observation of increased interfacial area at $t/\tau_e=15$ and $25$ time instance for the finer grid resolutions, $k_{\max} \eta_{KH} = 60$ and $120$, a closer qualitative comparison at the center $x-y$ plane for the proposed converged grid resolutions are presented in Figure~\ref{fig:visualizations_slices_cases_E}. Volume fraction contours over the turbulent flow field, represented by contours of enstrophy, are compared for the grid resolutions of $k_{\max} \eta_{KH} = 60$ and $120$ at three time instances $t/\tau_e = 10$, $15$, and $25$, where the total interfacial area is approximately equal between both resolutions at $t/\tau_e = 10$, higher for $k_{\max} \eta_{KH} = 120$ at $t/\tau_e = 15$, and higher for $k_{\max} \eta_{KH} = 60$ at $t/\tau_e = 25$. Overall, the size of the smallest bubbles/droplets is similar between both resolutions. For both time instances, it seems that the number density of small bubbles/droplets is initially higher for the finest grid resolution, being more pronounced at $t/\tau_e=15$. Nevertheless, this behavior reverses at $t/\tau_e=25$, indicating that the intermittent process occurs for both grid resolutions. 
%Although the proposed resolution of $k_{\max} \eta_{KH} = 60$ may not capture all sub-Kolmogorov-Hinze scales, the interfacial area contributions by those scales is not significant compared to larger scales as observed in the Figure~\ref{fig:area_contribution_cases_E}.

%\todo[inline]{i want to move the axis a little further away from the values}

\ifstrictfloat
\begin{figure}[H]
\else
\begin{figure}%[H]
\fi
\centering
\begin{tikzpicture}

% matrix of subfigures
\node (img11) {\includegraphics[width=0.35\textwidth]{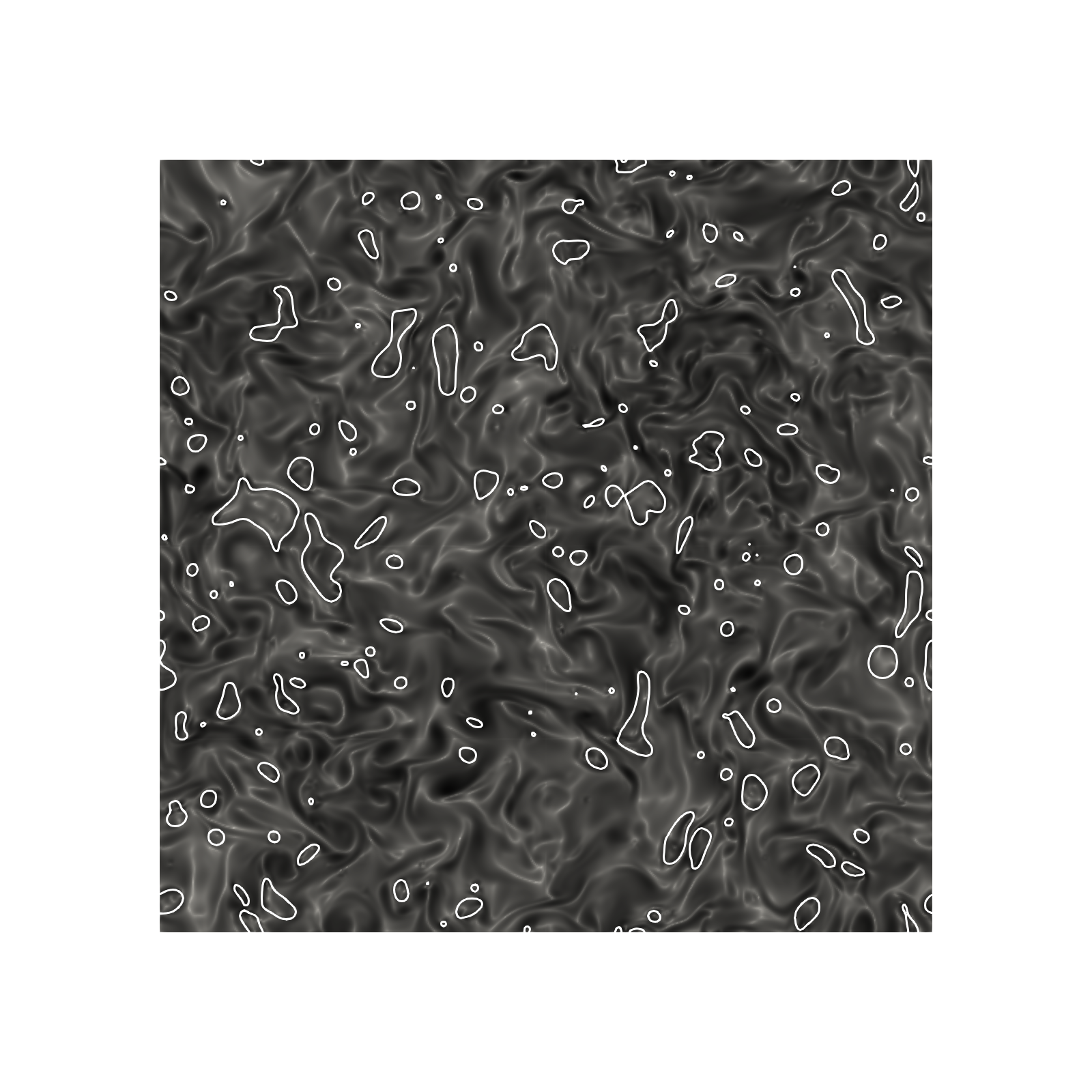}};
\node (img12) [right=0pt of img11] {\includegraphics[width=0.35\textwidth]{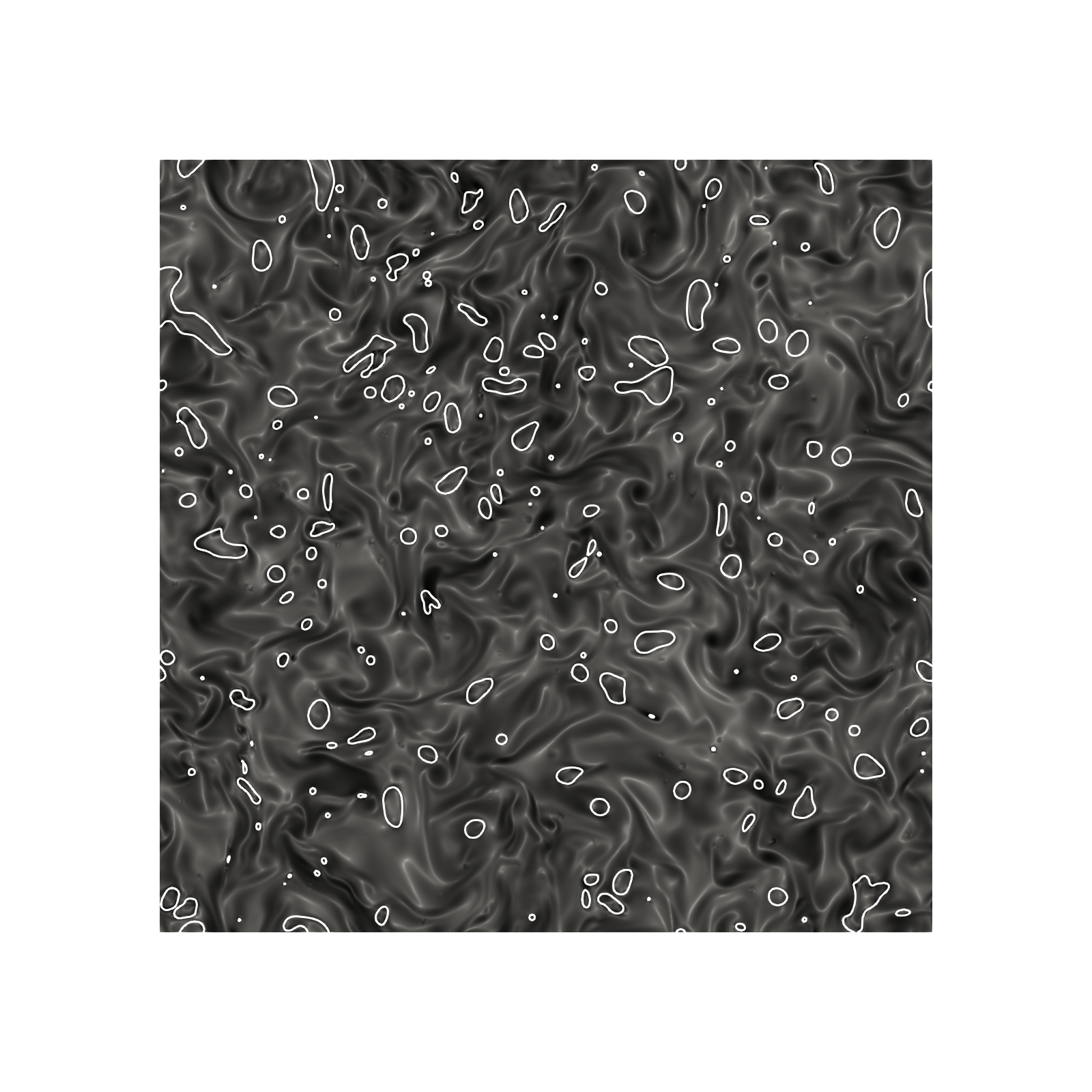}};
\node (img21) [below=0pt of img11] {\includegraphics[width=0.35\textwidth]{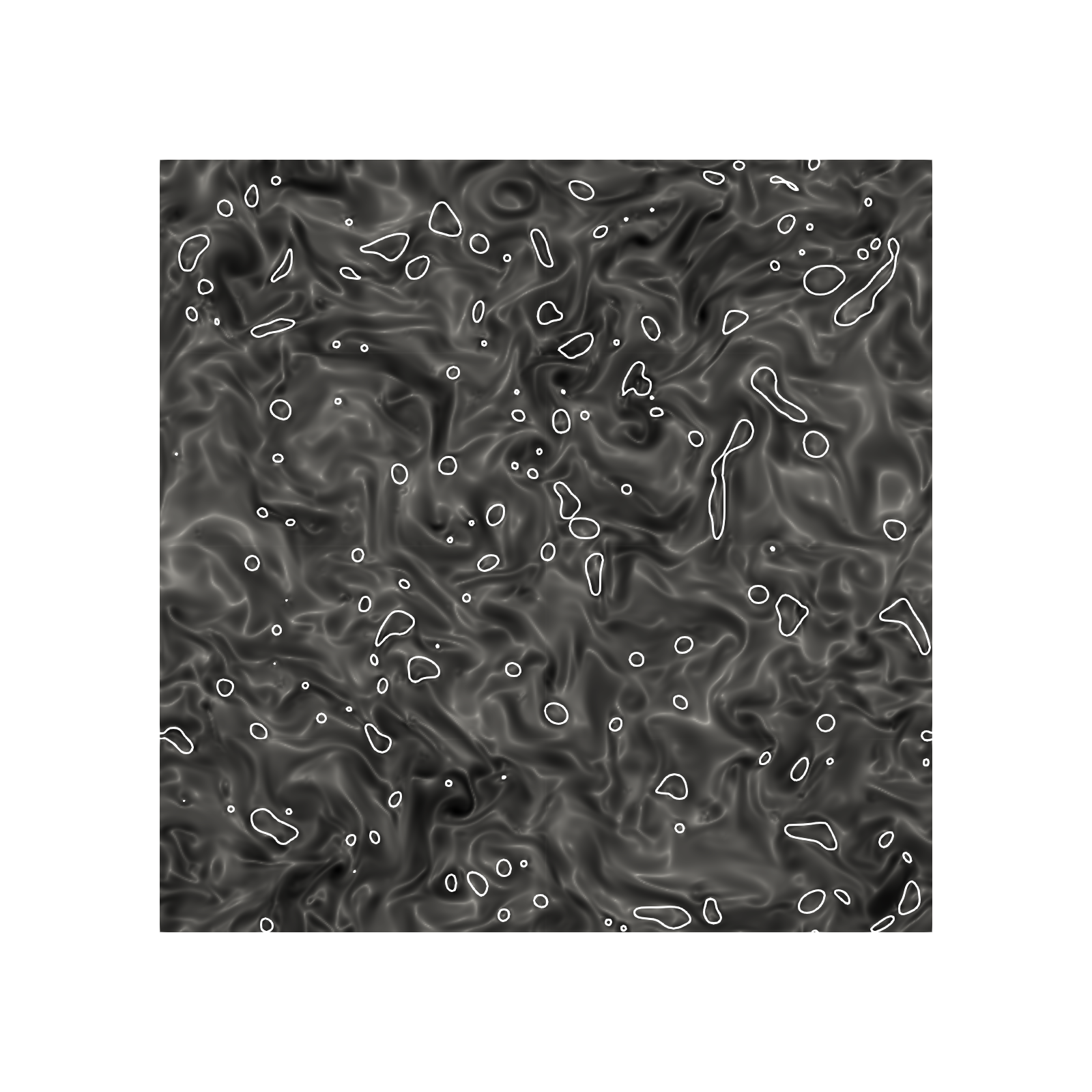}};
\node (img22) [below=0pt of img12] {\includegraphics[width=0.35\textwidth]{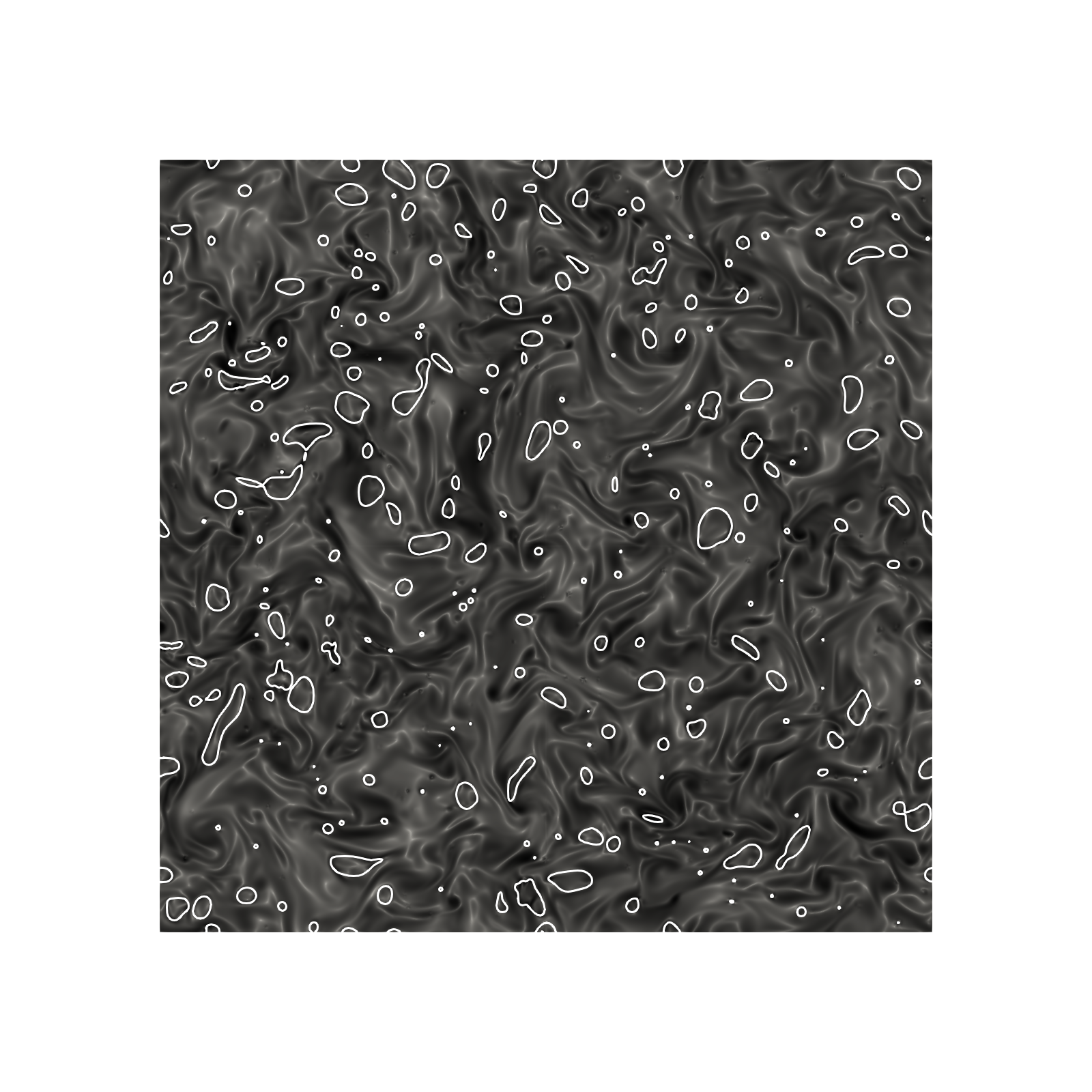}};
\node (img31) [below=0pt of img21] {\includegraphics[width=0.35\textwidth]{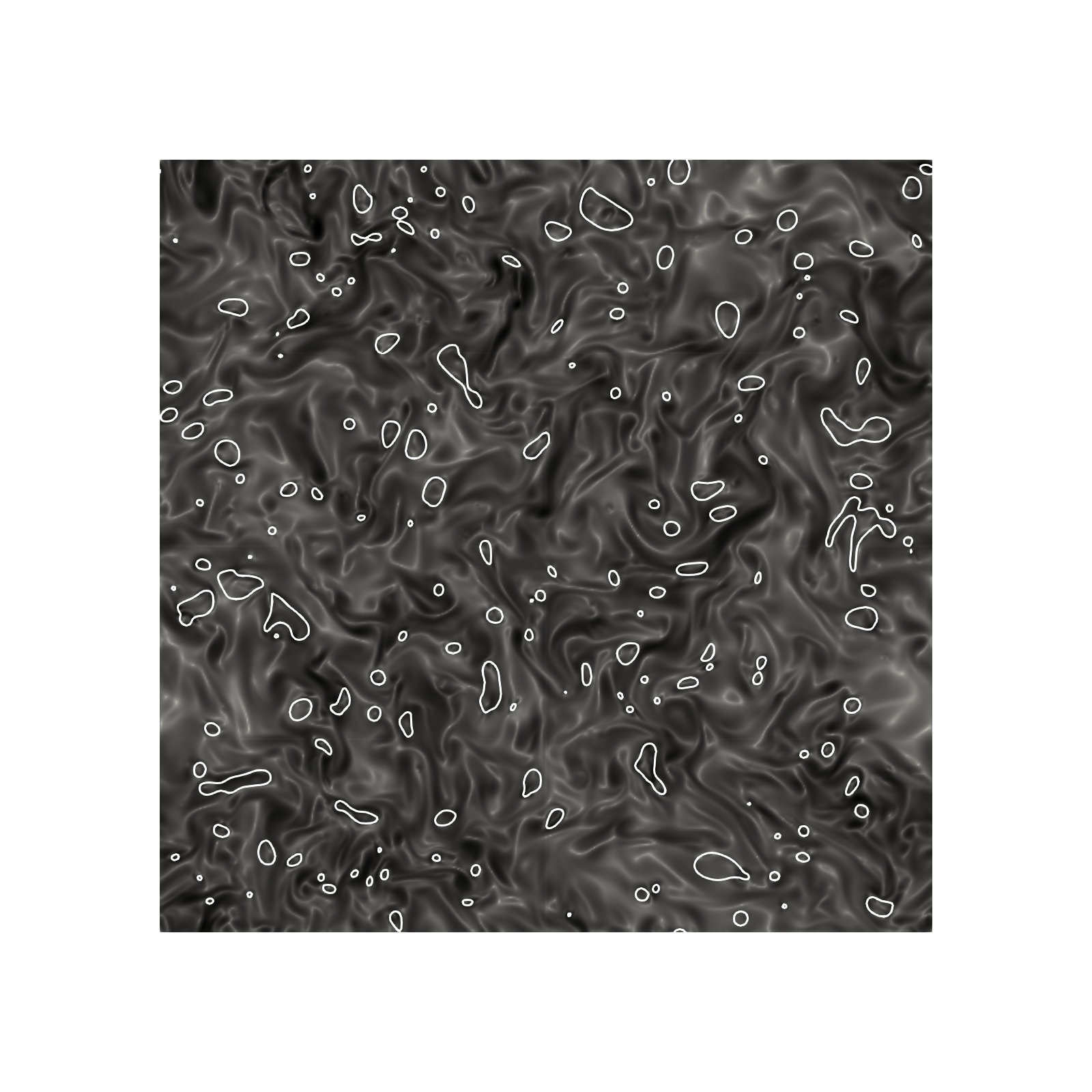}};
\node (img32) [below=0pt of img22] {\includegraphics[width=0.35\textwidth]{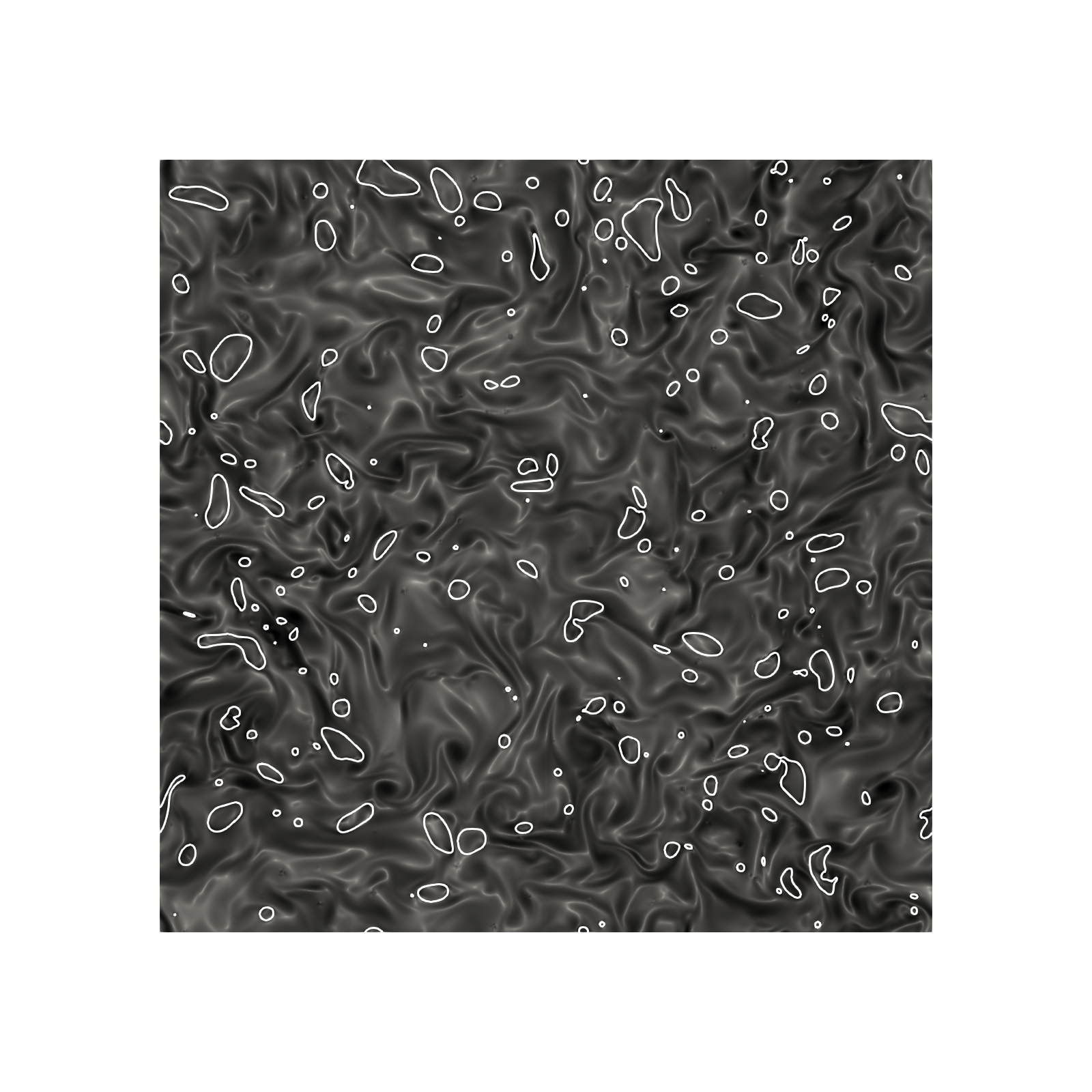}};

\begin{scope}[shift={(-7em,7em)}] % move the whole axis block closer to images
  % x-axis values
  \node (x1) at (2.6,0.5) {60};
  \node (x2) at (8.7,0.5) {120};
  
  % y-axis values
  \node (y1) at (-0.5,-2.6) {10};
  \node (y2) at (-0.5,-8.9) {15};
  \node (y3) at (-0.5,-14.8) {25};
  
  % axis titles
  \node at (5.7,1.15) {Resolution, $k_{\max}\eta_{KH} \; [-]$};
  \node[rotate=90] at (-1.2,-8.7) {Time, $t/\tau_e \; [-]$};
  
  % axis lines
  \draw[->, thick] (x1.north west) ++(-4,0) -- ($(x2.north east) + (3,0)$);
  \draw[->, thick] (y1.north west) ++(0,4) -- ($(y3.south west) + (0,-3)$);
\end{scope}

\end{tikzpicture}

\caption{Instantaneous volume fraction contours, $\phi = 0.5$, over center slice of enstrophy, $\Omega = \omega_i \omega_i$, of cases B, $Re_\lambda=87, We_\mathcal{L}=60$ for times $t/\tau_e = 10, 15$, and $25$ from top to bottom rows, respectively. Grid resolution values highlighted at the top of each column.}
\label{fig:visualizations_slices_cases_E}
\end{figure}
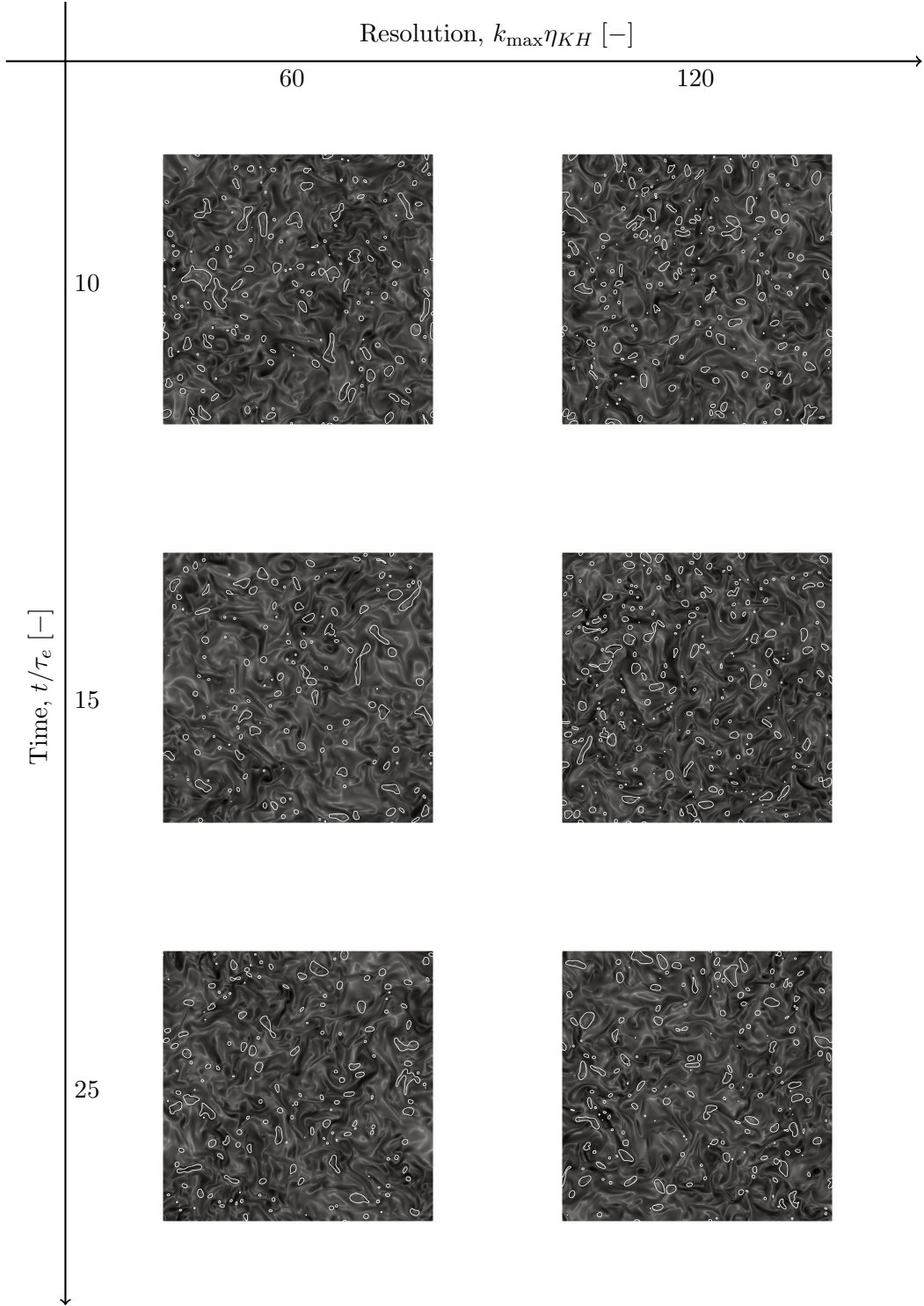

The Figure~\ref{fig:area_diss_time_cases_E} illustrates the fluctuations of the total interfacial area normalized by its mean value during the steady-state period in part (a), and the average energy dissipation rate normalized by its stationary value in part (b). One may observe that for grid resolutions of $k_{\max} \eta_{KH} \geq 60$, there is a significant correlation between both quantities. For instance, at $t/\tau_e \approx 15$, when there is an increase in the interfacial area for the grid resolution of $k_{\max} \eta_{KH} = 120$, there is also a significant increase in the average energy dissipation rate. Similarly, at $t/\tau_e \approx 25$, there is a significant increase in the interfacial area and the average energy dissipation rate for the grid resolution of $k_{\max} \eta_{KH} = 60$. 

\ifstrictfloat
\begin{figure}[H]
\else
\begin{figure}%[H]
\fi
    \centering
    \begin{subfigure}{0.45\linewidth}
    \includegraphics[width=\linewidth]{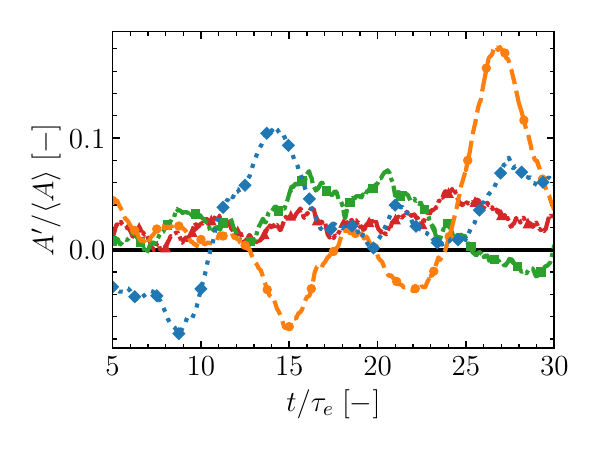}
    \caption{}
    \end{subfigure}
    \begin{subfigure}{0.475\linewidth}
    \includegraphics[width=\linewidth]{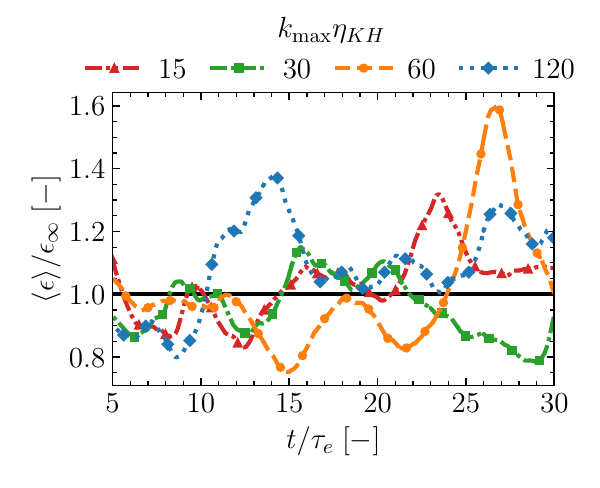}
    \caption{}
    \end{subfigure}
    \caption{Interfacial area fluctuation (a) and spatially averaged energy dissipation rate (b) as a function of time for different grid resolution for cases B, $Re_\lambda = 87$ and $We_\mathcal{L} = 60$.}
    \label{fig:area_diss_time_cases_E}
\end{figure}

The correlation between the dissipation rate and the interfacial area may be quantified using the Pearson's correlation coefficient, presented in Table~\ref{tab:vd_se_corr_cases_E}. For grid resolutions of $k_{\max} \eta_{KH} \ge 30$, the correlation coefficient is greater than 0.95. However, the magnitude of the interfacial area fluctuations for $k_{\max} \eta_{KH} = 30$ are not as significant as the ones for $k_{\max} \eta_{KH} \geq 60$. This is an indication that the proposed resolution of $k_{\max} \eta_{KH} = 60$ is able to capture a similar qualitative behavior in interfacial area fluctuations by the competing mechanisms of breakup and coalescence by the fluctuations in energy dissipation rate. 
%Although it is still not clear at which scales and by which events these fluctuations are affected by the turbulent structures, which are still of interest of ongoing efforts.

\ifstrictfloat
\begin{table}[H]%% placement specifier
\else
\begin{table}%[H]%% placement specifier
\fi
\centering%% For centre alignment of tabular.
\caption{Pearson's correlation coefficient between energy dissipation rate density and total interfacial area for cases B, $Re_\lambda = 87$ and $We_{\mathcal{L}}=60$.}\label{tab:vd_se_corr_cases_E}
\begin{tabular}{c c}%% Table column specifiers
  \hline
   $k_{\max} \eta_{KH}$ & $r$                  \\ \hline
   $15$                 & $0.7352058417284781$ \\
   $30$                 & $0.9516403599322205$ \\ 
   $60$                 & $0.9844842279585003$ \\ 
   $120$                & $0.9747155643410523$ \\ \hline
\end{tabular}
\end{table}

\bibliography{main}% Produces the bibliography via BibTeX.

\end{document}